\documentclass[a4paper,12pt]{article}

\usepackage{jheppub} 

\usepackage[T1]{fontenc} 

\usepackage{tikz}
\usepackage{import}
\usepackage{accents}
\usepackage{mathrsfs,amsmath,amssymb,slashed}
\usepackage{multirow,multicol}
\providecommand{\href}[2]{#2}

\usepackage[percent]{overpic}
\usepackage{slashed}
\usepackage{wrapfig}
\usepackage{tabu}
\usepackage{mathrsfs,amsmath,stackrel, amssymb,amsthm,amsfonts,tikz,graphicx,accents,hyperref}
\usepackage{dsfont,epiolmec, latexsym, stmaryrd, comment}
\usepackage{slashed,ccaption}
\usepackage{mathrsfs, calligra}
\usepackage{leftidx}
\usepackage{import}
\usepackage{multirow}
\usepackage{amsfonts}
\usepackage{tabularx}
\usetikzlibrary{intersections,calc}
\usepackage{tikz-3dplot}
\usepackage{ifthen}
\usetikzlibrary{arrows}
\usepackage{amsmath}
\usepackage{geometry}
\definecolor{rosy}{RGB}{230,235,252}
\definecolor{myframetitle}{RGB}{90,89,170}
\definecolor{myblocktitle}{RGB}{140,185,249}
\definecolor{mytitle}{RGB}{10,80,26}

\definecolor{darkgreen}{RGB}{27,130,45}
\definecolor{darkblue}{rgb}{0,0,0.3}
\definecolor{darkred}{rgb}{0.7,0,0}

\definecolor{light gray}{RGB}{220,220,220}
\definecolor{dark purple}{RGB}{108,0,217}
\definecolor{pink}{RGB}{190,20,100}
\definecolor{orange}{RGB}{193,63,0}
\definecolor{green}{RGB}{11,98,17}
\definecolor{darkpink}{RGB}{153,0,76}
\definecolor{bluegreen}{RGB}{0,102,102}
\definecolor{greenlagan}{RGB}{0,102,0}
\definecolor{redgreen}{RGB}{102,102,0}
\definecolor{Redgreen}{RGB}{153,76,0}

\usepackage{array}

\DeclareFontFamily{OT1}{rsfs}{}

\DeclareFontShape{OT1}{rsfs}{m}{n}{ <-7> rsfs5 <7-10> rsfs7 <10->rsfs10}{} 

\DeclareMathAlphabet{\mycal}{OT1}{rsfs}{m}{n}

\newcommand{\bmst}{{$\mathfrak{bms}_3$}}
\newcommand{\bmsf}{{$\mathfrak{bms}_4$}}
\newcommand{\hbmst}{$\widehat{\mathfrak{bms}}_3$}
\newcommand{\hbmsf}{$\widehat{\mathfrak{bms}}_4$}

\newcommand{\wfab}{${\mathcal{W}(a,b;\bar{a},\bar{b})}$}

\newcommand{\be}{\begin{equation}}
\newcommand{\ee}{\end{equation}}

\tdplotsetmaincoords{48}{112}
\makeatletter \@addtoreset{equation}{section}

\newcommand\hnote[1]{\textcolor{magenta}{\bf [Hamid:\,#1]}}

\newcommand\snote[1]{\textcolor{orange}{\bf [Sh-J:\,#1]}}
\preprint{IPM/P-2019/003}

\title{\centerline{\boldmath \emph{BMS$_4$ Algebra, Its Stability and Deformations}}}

\author[a]{H. R. Safari}
\author[a]{and M. M. Sheikh-Jabbari}

\affiliation{$^a$ School of Physics, Institute for Research in Fundamental
Sciences (IPM),\\ P.O.Box 19395-5531, Tehran, Iran}

\emailAdd{hrsafari@ipm.ir}
\emailAdd{jabbari@theory.ipm.ac.ir}

\abstract{
We continue analysis of \cite{Parsa:2018kys} and study rigidity and stability of the \bmsf\ algebra and its centrally extended version $\widehat{\mathfrak{bms}}_{4}$. We construct and classify the family of algebras which appear as deformations of \bmsf\ and in general find the four-parameter family of algebras \wfab\ as a result of the stabilization analysis, where  \bmsf$={\mathcal{W}}(-1/2,-1/2;-1/2,-1/2)$.
We then study the \wfab\ algebra, its maximal finite subgroups and stability for different values of the four parameters. We prove stability of the \wfab\ family of algebras for generic values of the parameters. For special cases of $(a,b)=(\bar{a},\bar{b})=(0,0)$ and $(a,b)=(0,-1), (\bar{a},\bar{b})=(0,0)$ the algebra can be deformed.
In particular we show that centrally extended $\mathcal{W}(0,-1;0,0)$ algebra can be deformed to an algebra which has three copies of Virasoro as a subalgebra. We briefly discuss these deformed algebras as asymptotic symmetry algebras and the physical meaning of the stabilization and implications of our result.
}
\begin{document}
\maketitle

\section{Introduction and motivations}

Motivated by a possible resolution to black hole information paradox and also by a rederivation and reinterpretation of soft theorems, studying algebras of ``soft charges'' has attracted a lot of attention, see \cite{Hawking:2016msc,Hawking:2016sgy, Strominger:2017zoo} and references therein or their citations list. Soft charges are associated with a specific subsector of gauge or diffeomorphism transformations which are singled out by an appropriate falloff behavior or boundary conditions. As such, the states carrying these charges all have the same energy. That is, turning on these charges do not alter the energy of the state, hence justifying the name soft charge. 

Being associated with (continuous) local gauge symmetries, there are generically infinite (but countable) number of soft charges and they form an infinite dimensional algebra. Prime examples of such algebras are the $\mathfrak{bms}_3$, $\mathfrak{bms}_4$ and two copies of Virasoro algebras, respectively studied in \cite{Ashtekar:1996cd, Oblak:2016eij}, \cite{Bondi:1962px, Sachs:1962zza, Sachs:1962wk, Barnich:2006av, Barnich:2009se, Barnich:2011mi, Duval:2014uva, Barnich:2017ubf, troessaert2018bms4} and \cite{Brown:1986nw, Ashtekar:1984zz}, and associated with asymptotic symmetry algebras of nontrivial diffeomorphisms on 3d and 4d flat spaces and on AdS$_3$. One may analyze charges and symmetry algebras of near-horizon nontrivial diffeomorphisms leading to $u(1)$ Kac-Moody algebra, Heisenberg algebra or $\mathfrak{sl}(2,\mathbb{R})$ current algebras in the context of 3d gravity theories \cite{Compere:2013bya, Afshar:2016wfy, Afshar:2016kjj, Afshar:2017okz, Grumiller:2016pqb, Grumiller:2017sjh}, or BMS-type or Heisenberg algebras in higher dimensional cases \cite{donnay:2015abr, NH-symmetry, grumiller:2018scv}. Analysis of nontrivial gauge transformations for the Maxwell theory leads to an infinite dimensional Abelian algebra \cite{kapec2014asymptotic} while addition of magnetic soft charges leads to an infinite dimensional Heisenberg algebra \cite{Hosseinzadeh:2018dkh, strominger:2015bla, campiglia:2016hvg}. Similar analysis may be carried out for $p$-form gauge theories and associated soft charges \cite{Afshar:2018apx, Francia:2018jtb}.

Deformation theory of Lie algebras has been introduced in 1960s  \cite{gerstenhaber1964deformation,gerstenhaber1966deformation, gerstenhaber1968deformation, gerstenhaber1974deformation, nijenhuis1967deformations} and immediately was applied by physicists to study important Lie algebras in physics \cite{levy1967deformation}. The idea is to analyze possible deformations one can make in structure constants of a given Lie algebra. Some of such deformations may just be a change of the basis which are called trivial deformations. There could be nontrivial deformations which deform the algebra into another algebra with the same number of generators. If an algebra does not admit any nontrivial deformation, it is called to be rigid or stable. For finite dimensional Lie algebras it has been proven that (Whitehead and Hochschild-Serre factorization theorems) any semi-simple Lie algebra is stable \cite{Whitehead-1, Whitehead-2,H-S-factorization-theorem}. These theorems relate the stability of the algebra to its second adjoint cohomology, see \cite{Parsa:2018kys} for a short review and summary. The deformation and stabilization of an algebra is inverse of the contraction procedure first introduced by In\"on\"u and Wigner \cite{Inonu:1953sp}. As a well-known example,   one may show that the Lorentz algebra may be contracted to Galilean algebra and conversely the Galilean algebra may be stabilized into the Lorentz algebra \cite{mendes1994deformations}. For a more recent analysis on stabilization of symmetry algebras with a rotation subgroup (``kinematical algebras'') see
\cite{Figueroa-OFarrill:1989wmj,Chryssomalakos:2004gk, Figueroa-OFarrill:2017sfs, Figueroa-OFarrill:2017ycu, Figueroa-OFarrill:2017tcy, Andrzejewski:2018gmz, Figueroa-OFarrill:2018ygf, Figueroa-OFarrill:2018ilb}.

The Hochschild-Serre factorization theorem, however, does not apply to infinite dimensional algebras, like the asymptotic symmetry algebras discussed above. In the absence of general theorems, stability of these algebras has been studied in case-by-case basis, e.g. see \cite{Fialowski:2001me, fialowski2012formal, gao2008derivations, gao2011low, Ecker:2017sen,Ecker:2018iqn}. In particular, in our previous paper \cite{Parsa:2018kys}, we studied stability of $\mathfrak{bms}_3$ and its centrally extended versions $\widehat{\mathfrak{bms}_3}$. Besides recovering cases where the algebra could be deformed into two Virasoro algebras,\footnote{It was shown in  \cite{Barnich:2012rz} that \hbmst\ may be obtained as an In\"on\"u contraction of two Virasoro algebras.} we found that $\mathfrak{bms}_3$ can be deformed into a two-parameter family of $W(a,b)$ algebras (which were first introduced in \cite{gao2011low}). This provides an explicit example of evasion of the Hochschild-Serre factorization for infinite dimensional algebras. We also analyzed rigidity of $W(a,b)$ algebras and showed that for generic values of $a,b$ parameters this family of algebras are stable.  
Furthermore, we studied how the deformation of the algebra interplays with admission of central extensions.

In this work we focus on the $\mathfrak{bms}_4$ algebra and study its deformations and stability. The ``original'' \bmsf\ algebra introduced in \cite{Bondi:1962px, Sachs:1962zza} is a semi-direct sum of Lorentz algebra with Abelian ideal spanned by supertranslations 
\begin{equation*}
    (\mathfrak{bms}_{4})_{\text{old}}=\text{Lorentz}\,\inplus\,\text{Supertranslations},
\end{equation*}
which has $4d$ Poincar\'e a subalgebra. Barnich and Troessaert in \cite{Barnich:2009se, Barnich:2011ct} suggested the Lorentz part of the $\mathfrak{bms}_4$ might be replaced by a larger (infinite dimensional) algebra called superrotations. In this work, as it is common in the recent literature,  we use \bmsf\ to denote this extended version. \bmsf\ is hence semi-direct sum of superrotations and supertranslations as
\begin{equation*}
   \mathfrak{bms}_{4}= \text{Superrotations}\,\inplus\,\text{Supertranslations}.
\end{equation*}
Later, they also classified its central extensions $\widehat{\mathfrak{bms}_4}$ in \cite{Barnich:2011ct,Barnich:2017ubf}.

There are many physical or mathematical motivations to carry out the stability analysis of the \bmsf. To state our main motivation, let us review some facts:
\begin{itemize}
    \item[(1)] As mentioned, the asymptotic symmetries of AdS$_3$ space is two Virasoro algebra at Brown-Henneaux central charge \cite{Brown:1986nw}. The seminal Brown-Henneaux analysis was a precursor of the celebrated AdS/CFT. 
    \item[(2)] This algebra upon the In\"on\"u contraction goes over to the \hbmst\ \cite{Barnich:2012rz}, which is asymptotic symmetry group of 3d flat space. This contraction is geometrically the large AdS radius $\ell$ limit under which the AdS$_3$ goes over to 3d flat space.
    \item[(3)] Under a similar large radius limit, geometrically,   AdS$_d$ space yields a $d$ dimensional flat Minkowski space for any $d$. 
    \item[(4)] It has been  argued that the asymptotic symmetry group of AdS$_d,\ d>3$ is nothing more than isometries of the spacetime $\mathfrak{so}(d-1,2)$ \cite{Ashtekar:1984zz, Henneaux:1985ey, henneaux1985asymptotically,  Ashtekar:1999jx}.
    \item[(5)] Asymptotic symmetry algebra analysis depends very much on the choice of boundary falloff behavior on metric fluctuations and there could always be a question whether the results mentioned in item (4) above could some how be evaded by a more relaxed boundary condition.
    \item[(6)] The asymptotic symmetry group of 4d flat space is known to be the infinite dimensional \bmsf\ algebra.\footnote{
Note that the notion of BMS algebra, which includes superrotations plus supertranslations, does not seemingly exist in dimensions higher than 4  \cite{Kapec:2015vwa,Hollands:2016oma}.}
    \end{itemize}
Therefore, it is natural to wonder if the \bmsf\ may come from contraction of an infinite dimensional ``asymptotic symmetry algebra of AdS$_4$''. In this work we  confirm these earlier results, in the sense that we show \bmsf\ algebra cannot be deformed into an algebra which has $\mathfrak{so}(3,2)$ as its subalgebra.
Our algebraic analysis and results has the advantage that it roles out possibility of existence of falloff conditions (for metric fluctuations) which may allow for a bigger symmetry algebra than $\mathfrak{so}(3,2)$ for AdS$_4$ asymptotic symmetry algebra.

As another motivation, we note that the asymptotic symmetry algebras has been argued to be relevant for formulation of holographic dual field theories, e.g. see \cite{Barnich:2010eb}. As argued in \cite{Parsa:2018kys} the deformation of asymptotic symmetry algebras may be attributed to the holographic renormalization of the conformal weight (scaling dimension) of  the operators. This analysis may also shed more light on field theory dual to gravity on 4d flat space. We will make more comments on this in the discussion section.

The full stability analysis of \bmsf\ algebra and its centrally extended version \hbmsf\ which we carry out here, reveals that \bmsf\ algebra can be deformed into a four parameter family of algebras \wfab . This algebra is an analogue of the $W(a,b)$ algebras obtained in \bmst\ stability analysis \cite{Parsa:2018kys}. We then study stability of the \wfab\ family of  algebras and show that they are generically stable. We also extend this analysis to the centrally extended versions of these 
algebras and classify all possible central extensions \wfab\ algebras can admit, for generic values of the four parameters as well as in the special points.

\paragraph{Organization of the paper.} In section \ref{sec:2}, we review and introduce \bmsf, its central extension $\widehat{\mathfrak{bms}}_4$ and its global subalgebra (4d Poincar\'e algebra). In section \ref{sec:3}, we study all possible infinitesimal deformations of \bmsf\ algebra. We show that this algebra can only be deformed to \wfab. In section \ref{sec:4}, we analyze the most general formal (finite) deformations of \bmsf\ and study integrability of the infinitesimal deformations. This section contains our main result on \bmsf, as we prove that \bmsf\ can only be deformed into \wfab. In section \ref{sec:5},  we study \wfab\ algebras, their subalgebras, deformations and stability and prove a theorem that \wfab\ family of algebras are stable for generic values of the four parameters. In section \ref{sec:6},  we repeat analysis of previous sections considering central extensions. In section \ref{sec:7}, we present algebraic cohomology arguments, based on Hochschild-Serre spectral sequence \cite{H-S-factorization-theorem, fuks2012cohomology}, for the \bmsf\ and \wfab\ algebras. In this way we provide a cohomological basis for our explicit computations of previous sections. Section \ref{sec:8} is devoted to summary of results and discussions. In a couple of appendices we have gathered some more technical analysis. In appendix \ref{appendix-A} we have reviewed the fact that generators of \bmsf\ or \wfab\ algebras may be viewed as functions on an $S^2$ and analyze implications of this on the index structure of possible deformations. In appendix \ref{appendix-B}, for completeness we have reviewed some basic facts of algebra cohomologies and the Hochschild-Serre spectral sequence.

\paragraph{Notation.} We adopt the same notation as \cite{Parsa:2018kys} for the algebras; for algebras we generically use ``mathfrak'' fonts, like $\mathfrak{witt}$, \bmst,\ \bmsf\ and $\mathfrak{KM}_{\mathfrak{u}(1)}$ ($u(1)$ Kac-Moody algebra). We will also be dealing with two and four parameter algebras, $W(a,b)$ and ${\cal W}(a,b;\bar{a},\bar{b})$, where in our conventions, \bmst$=W(0,-1)$, $\mathfrak{KM}_{\mathfrak{u}(1)}=W(0,0)$ and \bmsf$={\cal W}(-1/2,-1/2;-1/2,-1/2)$. 
The centrally extended version of an algebra $\mathfrak{g}$ will be denoted by $\hat{\mathfrak{g}}$, e.g. Virasoro algebra $\mathfrak{vir}=\widehat{\mathfrak{witt}}$.  We will be using ``$W(a,b)$ family'' of algebras (of $W(a,b)$ family, in short), to denote set of algebras for different values of the $a,b$ parameters and similarly for ${\cal W}(a,b;\bar{a},\bar{b})$ family. 
 
\section{Introduction to \texorpdfstring{$\mathfrak{bms}_4$}{BMS4}  algebra}\label{sec:2}

{In this section we  review the structure of  asymptotic symmetry algebras appearing in the context of 4d gravity.}
Depending on the asymptotic behavior of the metric and the chosen boundary falloff conditions one can get different asymptotic symmetry algebras. 

\subsection {{4d} flat space asymptotic symmetry algebra}

The centerless asymptotic symmetry  algebra of 4d flat spacetime is $\mathfrak{bms}_{4}$:
 \begin{equation} 
\begin{split}
 & [\mathcal{L}_{m},\mathcal{L}_{n}]=(m-n)\mathcal{L}_{m+n}, \\
 & [\bar{\mathcal{L}}_{m},\bar{\mathcal{L}}_{n}]=(m-n)\bar{{\mathcal L}}_{m+n},\\
 &[{\mathcal{L}}_{m},\bar{\mathcal{L}}_{n}]=0,\\
 &[\mathcal{L}_{m},T_{p,q}]=(\frac{m+1}{2}-p)T_{p+m,q},\\
 &[\bar{\mathcal{L}}_{n},T_{p,q}]=(\frac{n+1}{2}-q)T_{p,q+n},\\
 &[T_{p,q},T_{r,s}]=0,
\end{split}\label{bms4}
\end{equation}
where $m,n,p,q,r,s\in \mathbb{Z}$ and it is defined over the field of real numbers $\mathbb{R}$. {The $\mathfrak{bms}_{4}$ is an infinite dimensional algebra with countable basis} which is spanned by the generators $\mathcal{L}_m$, $\bar{\mathcal{L}}_{m}$ and $T_{p,q}$. The generators $\mathcal{L}_m$ and $\bar{\mathcal{L}}_{m}$ generate the direct sum of two Witt   subalgebra of  $\mathfrak{bms}_{4}$ and are usually called ``superrotations''.  $T_{p,q}$, the ``supertranslations,'' construct an adjoint representation of the direct sum of two Witt algebras and form the ideal part of $\mathfrak{bms}_{4}$.  Eq.\eqref{bms4} makes it clear that $\mathfrak{bms}_{4}$ has a semi-direct sum structure:
\begin{equation} \label{bms=witt+ideal}
\mathfrak{bms}_{4}= \big(\mathfrak{witt}\oplus\mathfrak{witt}\big)\inplus_{ad}\mathfrak{T}_{ab},
\end{equation}
where the subscript $ab$ is to emphasize $T_{p,q}$ being abelian and $ad$ denotes the adjoint action.  

The global part, i.e. the maximal finite subalgebra, of $\mathfrak{bms}_{4}$ is 4d Poincar\'{e} algebra $\mathfrak{iso}(3,1)$ and is generated by ${\cal L}_0, {\cal L}_{\pm1}$ and $\bar{\cal L}_0, \bar{{\cal L}}_{\pm1}$ (which form Lorentz algebra $\mathfrak{so}(3,1)$) and 
$T_{0,0}, T_{0,1}, T_{1,0}, T_{1,1}$ as the translations. In the next subsection we will make the connection to the more usual basis for Poincar\'e algebra  explicit. 

The above $\mathfrak{bms}_{4}$ admits central extensions in ${\cal L}_n$, $\bar{\cal L}_n$ sectors \cite{Barnich:2011ct,Barnich:2011mi}. The centrally extended algebra which will be denoted by $\widehat{\mathfrak{bms}}_{4}$ is
 \begin{equation} 
\begin{split}
 & [\mathcal{L}_{m},\mathcal{L}_{n}]=(m-n)\mathcal{L}_{m+n}+\frac{C_{\mathcal{L}}}{12}(m^{3}-m)\delta_{m+n,0}, \\
 & [\bar{\mathcal{L}}_{m},\bar{\mathcal{L}}_{n}]=(m-n)\bar{\mathcal{L}}_{m+n}+\frac{C_{\bar{\mathcal{L}}}}{12}(m^{3}-m)\delta_{m+n,0},\\
 &[{\mathcal{L}}_{m},\bar{\mathcal{L}}_{n}]=0,\\
 &[\mathcal{L}_{m},T_{p,q}]=(\frac{m+1}{2}-p)T_{p+m,q},\\
 &[\bar{\mathcal{L}}_{n},T_{p,q}]=(\frac{n+1}{2}-q)T_{p,q+n},\\
 &[T_{p,q},T_{r,s}]=0,
\end{split}\label{BMS-centrally-extended}
\end{equation}
in which $C_{\mathcal{L}}$ and $C_{\bar{\mathcal{L}}}$ are called central charges. One may readily see that the central terms, which vanish for $m=0,\pm 1$, do not appear in the global part of the algebra. Therefore, global part of $\widehat{\mathfrak{bms}}_{4}$ is also 4d Poincar\'{e}. We note that the second real cohomology of the $\mathfrak{bms}_{4}$ algebra, ${\cal H}^2(\mathfrak{bms}_{4}; \mathbb{R})$, which  classifies (global) central extensions of the algebra does not allow for any other central extension \cite{Barnich:2011ct}, other than $C_{\mathcal{L}}, C_{\bar{\mathcal{L}}}$.

\subsection {More on global part of the \texorpdfstring{$\mathfrak{bms}_4$}{BMS4} algebra}

As mentioned the $\mathfrak{bms}_4$ algebra which is the asymptotic symmetry algebra of 4d flat space, should contain Poincar\'e algebra which is the isometry algebra of the flat space. Generators of the Poincar\'e algebra is usually written in the 4d tensorial basis, $J^{\mu \nu}$, the generator of Lorentz algebra, and $P^{\mu}$, the generator of translations, as it discussed in \cite{weinberg1995quantum} 
\begin{equation} \label{poincareJ-P}
\begin{split}
 & [J^{\mu \nu},J^{\rho \sigma}]=i(\eta^{\mu \rho} J^{\nu \sigma}+\eta^{\sigma \mu}J^{\rho \nu}-\eta^{\nu \rho}J^{\mu \sigma}-\eta^{\sigma \nu}J^{\rho \mu}), \\
  &[J^{\mu \nu},P^{\rho}]=i(\eta^{\rho  \mu}P^{\nu}-\eta^{\rho \nu}P^{\mu}), \\
 &[P^{\mu},P^{\nu}]=0,
\end{split}
\end{equation}
where $\mu, \nu=0,1,2,3$ and $\eta_{\mu\nu}=diag(-,+,+,+)$ is the Minkowski metric. To relate the above algebra to the global part of $\mathfrak{bms}_4$ one should decompose the Lorentz part into $\mathfrak{sl}(2)\oplus\mathfrak{sl}(2)$ basis:
\begin{equation*}
\begin{split}
    & \mathcal{L}_{\pm}\equiv iR^{1}\pm R^{2},\\
     &\bar{\mathcal{L}}_{\pm}\equiv iL^{1}\pm L^{2},\\
     & \mathcal{L}_{0} \equiv R^{3},\,\,\,\,\,\ \bar{\mathcal{L}}_{0} \equiv L^{3},
\end{split}
\end{equation*}
where 
\begin{equation}
     {L}^{i}\equiv \frac{1}{2}(\mathcal{J}^{i}+i\mathcal{K}^{i}),\quad
     {R}^{i}\equiv \frac{1}{2}(\mathcal{J}^{i}-i\mathcal{K}^{i}), \qquad i=1,2,3.
\end{equation}
Here ${\cal J}^i, {\cal K}^i$ are generators of rotation and boost: 
\begin{equation*}
    \mathcal{J}^{i}:=\frac{1}{2}\epsilon^{i}_{\ jk} J^{jk},\,\,\,\,\,\ \mathcal{K}^{i}:=J^{0i},
\end{equation*}
where $\epsilon^{i}_{\ jk}$ is an antisymmetric quantity with $\epsilon^{1}_{\ 23}=+1$. In other words, the generators of 4d Lorentz algebra ${\cal J}^{\mu\nu}$ can be decomposed as $({\bf 1},{\bf 3})\oplus ({\bf 3},{\bf 1})$ of $\mathfrak{sl}(2)\oplus\mathfrak{sl}(2)$ algebra. One may then readily show that
\begin{equation} 
\begin{split}
 & [\mathcal{L}_{m},\mathcal{L}_{n}]=(m-n)\mathcal{L}_{m+n}, \\
  &[\mathcal{L}_{m},\bar{\mathcal{L}}_{n}]=0, \\
 &[\bar{\mathcal{L}}_{m},\bar{\mathcal{L}}_{n}]=(m-n)\bar{\mathcal{L}}_{m+n},
\end{split}\label{PoincareL-L}
\end{equation}
where $m,n=\pm 1,0$. 

The translation generators, which are Lorentz four-vectors $P_\mu$ can be decomposed into $({\bf 2},{\bf 2})$ of $\mathfrak{sl}(2)\oplus\mathfrak{sl}(2)$ algebra, i.e. $P_\mu$ are linear combinations of $T_{m,n}, m,n=0,1$:
\begin{equation}\label{basisP-T}
\begin{split}
 & P^{0}\equiv H=(T_{1,0}-T_{0,1}), \\
  &P^{3}=(T_{1,0}+T_{0,1}), \\
 &P^{1}=(-i)(T_{1,1}+T_{0,0}),\\
 & P^{2}=(T_{1,1}-T_{0,0}).
\end{split}
\end{equation}

\subsection{\texorpdfstring{AdS$_4$}{4AdS} isometry, \texorpdfstring{$\mathfrak{so}(3,2)$}{AdS4} algebra}

For our later use we also discuss the AdS$_4$ isometry algebra generated by $J^{ab}, a,b=-1,0,1,2,3$
\begin{equation} 
\begin{split}
 & [J^{a b},J^{c d}]=i(\eta^{a c} J^{b d}+\eta^{d a}J^{c b}-\eta^{b c}J^{a d}-\eta^{d b}J^{c a}), 
\end{split}\label{AdS-J}
\end{equation}
where $\eta^{ab}=(-1,-1,+1,+1,+1)$. $J^{ab}$ which is in ${\bf 15}$ representation of $\mathfrak{so}(3,2)$ may be decomposed in terms of $\mathfrak{sl}(2)\oplus\mathfrak{sl}(2)$ as $({\bf 1},{\bf 3})\oplus ({\bf 3},{\bf 1})\oplus ({\bf 3},{\bf 3})$.
The first two are just ${\cal L}_m, \bar{\cal L}_m, m=0,\pm 1$, and the last one may be denoted by $T_{m,n}, m,n=0,\pm 1$ with the commutation relations:
 \begin{equation}\label{AdS4}
\begin{split}
 & [\mathcal{L}_{m},\mathcal{L}_{n}]=(m-n)\mathcal{L}_{m+n}, \\
 & [\bar{\mathcal{L}}_{m},\bar{\mathcal{L}}_{n}]=(m-n)\bar{{\mathcal L}}_{m+n},\\
 &[{\mathcal{L}}_{m},\bar{\mathcal{L}}_{n}]=0,\\
 &[\mathcal{L}_{m},T_{p,q}]=(\frac{m+1}{2}-p)T_{p+m,q},\\
 &[\bar{\mathcal{L}}_{n},T_{p,q}]=(\frac{n+1}{2}-q)T_{p,q+n},\\
 &[T_{m,n},T_{p,q}]=\frac{1}{2}\bigg((q-n)\mathcal{L}_{m+p-1}+(p-m)\bar{\mathcal{L}}_{q+n-1}\bigg),
\end{split}
\end{equation}
where $m,n,p,q=0,\pm 1$. 

It is known that $\mathfrak{iso}(3,1)$ is not a rigid (stable) algebra and may be deformed into $\mathfrak{so}(3,2)$ or $\mathfrak{so}(4,1)$, which are stable \cite{levy1967deformation}. In the $\mathfrak{sl}(2)\oplus\mathfrak{sl}(2)$ notation adopted above, one may readily use the Hochschild-Serre factorization theorem to argue that only the ideal part of $\mathfrak{iso}(3,1)$, the $[T,T]$ commutator, can be deformed such that only ${\cal L}_m, \bar{\mathcal{L}}_m$ appears in the right-hand-side of the commutator. This can be manifestly seen in the last equation in \eqref{AdS4}.\footnote{Conversely, one may view $\mathfrak{iso}(3,1)$ algebra as the In\"on\"u-Wigner contraction of the $\mathfrak{so}(3,2)$ algebra. Geometrically, this contraction corresponds to a large radius limit of an AdS$_4$ space yielding a 4d flat space.} As reviewed in the introduction, these theorems do not apply to the infinite dimensional algebras and one cannot extend the above result which is about the global part $\mathfrak{bms}_4$ to the whole algebra. We will show in the rest of this work that there is no infinite dimensional algebra in the family of $\mathfrak{bms}_4$ deformations which has $\mathfrak{so}(3,2)$ as its global part.

\section{Deformations of \texorpdfstring{$\mathfrak{bms}_4$}{BMS4}  algebra}\label{sec:3}

In this section we consider deformations of  $\mathfrak{bms}_{4}$ defined in (\ref{bms4}).
As discussed the Hochschild-Serre factorization theorem is not applicable for infinite dimensional Lie algebras and working with them is more complicated than finite dimensional cases. Here, we first analyze possible deformations of $\mathfrak{bms}_{4}$ algebra by deforming  each commutation relation of $\mathfrak{bms}_{4}$ algebra separately. Of course one should check that in this way we do not miss any possible deformation which may involve more than one set of commutators. Finally, we study obstructions, which infinitesimal deformations yield formal deformations and  what are the rigid algebras obtained from deformations of $\mathfrak{bms}_{4}$.

\subsection{ Deformation in the two Witt subalgebras}\label{Witt-deformations-sec}

The Witt algebra is known to be rigid and hence there is no 2-cocycle which deforms $[\mathcal{L}_{m},\mathcal{L}_{n}]$ by coefficients of ${\cal L}_p$ \cite{fialowski2012formal, schlichenmaier2014elementary}. Similarly, we cannot deform $\mathfrak{witt}\oplus\mathfrak{witt}$ algebra \cite{Parsa:2018kys}. Therefore, in this sector the only option  is deforming $\mathfrak{witt}\oplus\mathfrak{witt}$ sector by coefficients of $T_{m,n}$ generators:\footnote{Here we are allowing for $d,\bar{d}$ to take arbitrary  values. However, as the discussions in the appendix \ref{appendix-A} indicates one could have fixed them by the requirement that the generators are functions on the $S^2$. This, however, does not affect our analysis and results in this subsection as all these deformations happen to be trivial and may be absorbed into redefinition of generators.} 
\begin{equation} \label{h-barh-H-deformation}
\begin{split}
    &[\mathcal{L}_{m},\mathcal{L}_{n}]=(m-n)\mathcal{L}_{m+n}+(m-n)\sum_{d,\bar{d}} h^{d,\bar{d}}(m,n)T_{m+n+d,\bar{d}},\\
    &[\bar{\mathcal{L}}_{m},\bar{\mathcal{L}}_{n}]=(m-n)\bar{\mathcal{L}}_{m+n}+(m-n)\sum_{d,\bar{d}} \bar{h}^{d,\bar{d}}(m,n)T_{d,m+n+\bar{d}},\\
    &[\mathcal{L}_{m},\bar{\mathcal{L}}_{n}]=\sum_{d,\bar{d}} H^{d,\bar{d}}(m,n)T_{m+d,n+\bar{d}},
\end{split}
\end{equation}
where $h,\bar{h}$ are symmetric and $H$ is arbitrary functions and $d$  and $\bar{d}$ are arbitrary numbers but we should note that the indices $d, \bar d$  should be equal with each other in all three relations.

The Jacobi $[\mathcal{L}_{m},[\mathcal{L}_{n},\mathcal{L}_{l}]]+cyclic\,\,\,permutation=0$ leads to 
\begin{equation} 
\begin{split}
    &\sum_{d,\bar{d}}\big((n-l)(m-n-l)h^{d,\bar{d}}(m,n+l)+(n-l)(\frac{m+1}{2}-n-l-d)h^{d,\bar{d}}(n,l)+\\
    &(l-m)(n-m-l)h^{d,\bar{d}}(n,m+l)+(l-m)(\frac{n+1}{2}-m-l-d)h^{d,\bar{d}}(l,m)+\\
    &(m-n)(l-m-n)h^{d,\bar{d}}(l,m+n)+(m-n)(\frac{l+1}{2}-m-n-d)h^{d,\bar{d}}(m,n)\big)T_{m+n+l+d,\bar{d}}=0,
\end{split}
\end{equation}
which its solution is $h^{d,\bar{d}}(m,n)=constant=h^{d,\bar{d}}$. The same relation can be obtained from the Jacobi $[\bar{\mathcal{L}}_{m},[\bar{\mathcal{L}}_{n},\bar{\mathcal{L}}_{l}]]+cyclic\,\,\,permutation=0$ for $\bar{h}^{d,\bar{d}}(m,n)$ which its solution is $\bar{h}(m,n)=constant=\bar{h}^{d,\bar{d}}$. 

The next two Jacobi identities to analyze are  $[\mathcal{L}_{m},[\mathcal{L}_{n},\bar{\mathcal{L}}_{l}]]+cyclic\,permutation=0$ and $[\bar{\mathcal{L}}_{m},[\bar{\mathcal{L}}_{n},\mathcal{L}_{l}]]+cyclic\,permutation=0$ which yield
\begin{align}
\label{H-h}   (\frac{m+1}{2}-n-d)H^{d,\bar{d}}(n,l)&-(\frac{n+1}{2}-m-{d})H^{d,\bar{d}}(m,l)-\cr
&-(m-n)\left(H^{d,\bar{d}}(m+n,l)-
(\frac{l+1}{2}-\bar{d})h^{d,\bar{d}}\right) =0,\\
\label{H-barh}   (\frac{n+1}{2}-m-\bar{d})H^{d,\bar{d}}(l,m)&-(\frac{m+1}{2}-n-\bar{d})H^{d,\bar{d}}(l,n)+\cr &+(m-n)\left(H^{d,\bar{d}}(l,m+n)+
    (\frac{l+1}{2}-{d})\bar{h}^{d,\bar{d}}\right)=0.
\end{align} 
Let us first consider the special case of $h^{d,\bar{d}}=\bar {h}^{d,\bar{d}}=0$. In this case one can easily see that 
\be\label{H0dd}
H_0^{d,\bar{d}}(m,n)=H^{d,\bar{d}}_0 (m+1-2d)(n+1-2\bar{d}),
\ee
where $H^{d,\bar{d}}_0$ is an arbitrary coefficient. 
Next, let us consider the generic case where $h^{d,\bar{d}},\bar {h}^{d,\bar{d}}\neq 0$. In this case the solution is of the form 
$$
H^{d,\bar{d}}(m,n)=H_0^{d,\bar{d}}(m,n)+\tilde{H}^{d,\bar{d}}(m,n),
$$
where $\tilde{H}^{d,\bar{d}}(m,n)$ is a solution to \eqref{H-h} and \eqref{H-barh} which vanishes as $h, \bar{h}=0$. The form of equations \eqref{H-h} and \eqref{H-barh} suggests that the most general solution is of the form $H(m,n)=a mn+ bm+c n+d$. The $mn$ term, however, can be absorbed in the ``homogeneous solution'' part $H_0^{d,\bar{d}}(m,n)$. 
Therefore, we consider the solution ansatz 
$\tilde{H}^{d,\bar{d}}(m,n)=Am+Bn+C$. Plugging this into \eqref{H-h} and \eqref{H-barh} yields
$A=\bar{h},\,\,\,B=-h,\,\,\,C=h(2\bar{d}-1)+\bar{h}(1-2d)$. To summarize, 
\begin{equation}
    H^{d,\bar{d}}(m,n)=H^{d,\bar{d}}_0 (m+1-2d)(n+1-2\bar{d})+\bar{h}(m+1-2d)-h(n+1-2\bar{d}).
\end{equation}

\paragraph{On triviality of these deformations.} One may examine whether the  $h,\bar h$ and $H(m,n)$ deformations are nontrivial or may be absorbed in the redefinition of generators. To this end let us consider redefined generators as
\begin{equation}\label{redefinitionXY}
\begin{split}
    &\Tilde{\mathcal{L}}_{m}\equiv \mathcal{L}_{m}+\sum_{d,\bar{d}}\ X^{d,\bar{d}}(m)T_{m+d,\bar{d}} ,\\
    &\Tilde{\bar{\mathcal{L}}}_{m}\equiv \bar{\mathcal{L}}_{m}+\sum_{d,\bar{d}}\ Y^{d,\bar{d}}(m)T_{d,m+\bar{d}},\\
    &\Tilde{T}_{m,n}\equiv T_{m,n},
\end{split}
\end{equation}
 where $X^{d,\bar{d}}(m)$ and $Y^{d,\bar{d}}(m)$ are  functions to be determined upon requirement of removing $h^{d,\bar{d}},{\bar h}^{d,\bar{d}}$ and $H^{d,\bar{d}}$ terms in  \eqref{h-barh-H-deformation}. Removal of $h,\bar{h}$, i.e.  
requiring $[\Tilde{\mathcal{L}}_{m},\Tilde{\mathcal{L}}_{n}]=(m-n)\Tilde{\mathcal{L}}_{m+n}$ and performing the same analysis for $\Tilde{\bar{\mathcal{L}}}_{m}$ yields $X^{d,\bar{d}}(m)=A(m+1-2d)-2h^{d,\bar{d}},\ Y^{d,\bar{d}}(m)=\bar{A}(m+1-2\bar{d})- 2\bar{h}^{d,\bar{d}}$. 
Requiring $[\Tilde{\mathcal{L}}_{m},\bar{\tilde{\mathcal{L}}}_{n}]=0$, yields $A-\bar{A}=2H_0^{d,\bar{d}}$. One may take $A=-\bar{A}=H_0^{d,\bar{d}}$ and hence
\be\label{Redef-XY}
X^{d,\bar{d}}(m)=H_0^{d,\bar{d}}(m+1-2d)-2h^{d,\bar{d}},\qquad Y^{d,\bar{d}}(m)=-H_0^{d,\bar{d}}(m+1-2\bar{d})- 2\bar{h}^{d,\bar{d}}
\ee
would remove the deformations. Therefore, the deformations in \eqref{h-barh-H-deformation} are all trivial.


\subsection{ Deformation of  \texorpdfstring{$[\mathcal{L},T]$}{LT} commutators}

The deformations in this sector could be with coefficients of $T_{m,n}$ or ${\cal L}_m$. We consider these two cases separately. 

\paragraph{With coefficients in \texorpdfstring{$T$}{T}.}
Consider the deformations of commutator of superrotations and supertranslations which is the fourth line in (\ref{bms4}) without changing other commutators. To this end as in the previous subsection we add a 2-cocycle function:
\begin{equation} 
 [\mathcal{L}_{m},T_{p,q}]=(\frac{m+1}{2}-p)T_{p+m,q}+ K(m,p)T_{p+m,q},\label{LT-Tdeformation}
\end{equation}
We have fixed the first index of $T$ on the right-hand-side to be $m+p$, see appendix \ref{appendix-A} for further discussions. Here, $K(m,n)$ is an unknown function to be determined through closure of algebra requirements.

To find the explicit form of function $K(m,n)$, there are two Jacobi identities to check. The first one is  $[\mathcal{L}_{m},[\mathcal{L}_{n},T_{p,q}]]+{cyclic\ permutations}=0$, which to first order in the deformation parameter yields
\begin{multline}
(\frac{n+1}{2} - p) K(m, p + n) + ( \frac{m+1}{2}-p-n) K(n, p) + (p-\frac{m+1}{2}) K(
   n, p+m) +\\  +
  (p+m - \frac{n+1}{2}) K(m,p)+ (n-m) K(m+n,p)=0.\label{eq-K}
\end{multline}
For $p,m=0$ we get
\begin{equation} 
 (\frac{n+1}{2}) (K(0, n) - K(0, 0))=0,\nonumber
\end{equation}
and hence
\begin{equation} 
  K(0, n)=\text{constant}.\label{first contraint on K}
\end{equation}
To solve \eqref{eq-K} we note that it is linear in $K$ and hence linear combination of any two solutions is also a solution. One may then check that 
\be\label{K-linear}
K(m,n)= \alpha+\beta m,
\ee
is a solution for any $\alpha,\beta$. This equation has solutions which involve higher powers of $m,n$. One may then examine a  degree $N$, i.e. $K(m,n)=\sum_{r=1}^N A_{rs} m^r n^{s}$ ansatz. At $N=2$ we obtain
a solution of the form
\begin{equation} 
 K(m,n)=\gamma m(\frac{m+1}{2}-n),\label{Ksolution}
\end{equation}
where $\gamma$ is an arbitrary constant and we have added the 1/2 factor for later convenience. This solution, however, is a trivial deformation as it can be absorbed in rescaling of $T_{m,n}$ generators:
\be\label{T-redef}
T_{m,n}\to \Tilde{T}_{m,n}=M(m) T_{m,n}
\ee
with $M(m)=1+\gamma m$. In general, one can show that the most general solution to \eqref{eq-K} is 
\be\label{K-redef}
K(m,n)=(\frac{m+1}{2}-n)\left(\frac{M(m+n)}{M(n)}-1\right),
\ee
which again can be absorbed in a redefinition of $T$ of the form \eqref{T-redef}. (Note that here $K(m,n)$ is to be viewed as an infinitesimal function. In section \ref{Integrability-Wab} we discuss finite deformations.\footnote{A similar pattern was also found in the 3d case, \emph{cf.} section 4.2 of \cite{Parsa:2018kys}.})

The other Jacobi to be checked is $[\bar{{\cal L}}_{m},[{\cal L}_{n},T_{p,q}]]+{cyclic\ permutations}=0$, which  does not yield a new constraint on $K(m,p)$. So the most general solutions of  \eqref{eq-K} are those we have derived. Deformations in  $[\bar{{\cal L}}_{n},T_{p,q}]$ can be analyzed in a similar manner, yielding similar results.

To summarize, the only non-trivial deformations are those generated by \eqref{K-linear} which yields $\mathcal{W}(a,b;\bar{a},\bar{b})$ algebra defined through  commutation relations,
 \begin{equation} 
\begin{split}
 & [\mathcal{L}_{m},\mathcal{L}_{n}]=(m-n)\mathcal{L}_{m+n}, \\
 & [\bar{\mathcal{L}}_{m},\bar{\mathcal{L}}_{n}]=(m-n)\bar{\mathcal{L}}_{m+n},\\
 &[{\mathcal{L}}_{m},\bar{\mathcal{L}}_{n}]=0,\\
 &[\mathcal{L}_{m},T_{p,q}]=-(p+bm+a)T_{p+m,q},\\
 &[\bar{\mathcal{L}}_{n},T_{p,q}]=-(q+\bar{b}n+\bar{a})T_{p,q+n},\\
 &[T_{p,q},T_{r,s}]=0.
\end{split}\label{W4-algebra}
\end{equation}
The above is a 4d extension of the $W(a,b)$ algebra which is a deformation of \bmst \cite{Parsa:2018kys}.

One may wonder if the index of $T$ generator appearing in the RHS of $[{\cal L}_n,T_{p,q}]$ is restricted to be $T_{n+p,q}$. As in the previous subsection, \emph{cf.} \eqref{h-barh-H-deformation},  the answer is no, at least as far as the Jacobi identity and the closure of the algebra is concerned. Explicitly, let us consider the following deformation,
\begin{equation} 
 [\mathcal{L}_{m},T_{p,q}]=(\frac{m+1}{2}-p)T_{p+m,q}+K(m,p)T_{p+m+d_0,q+\bar{d}_0},
\end{equation}
where $d_0,\bar{d}_0$ are two arbitrary constants.  The Jacobi identity then leads to
\begin{multline}\label{K-d}
(\frac{n+1}{2} - p) K(m, p + n) + ( \frac{m+1}{2}-p-n-d_0) K(n, p) + (p-\frac{m+1}{2}) K(
   n, p+m) +\\  +
  (p+m+d_0- \frac{n+1}{2}) K(m,p)+ (n-m) K(m+n,p)=0.
\end{multline}
It can be readily seen that for $d\neq 0$ the only solution to \eqref{K-d} is $K(m,n)=K=constant$. Nonetheless, this is trivial deformation, as it can be absorbed in the redefinition of $T_{m,n}$ as follows:
\[
\tilde{T}_{m,n}=\sum_{d} C_d T_{m+d,n}
\]
where $C_d$ are coefficients to be fixed upon request that $[{\cal L}_n,\tilde{T}_{p,q}]=(\frac{n+1}{2}-p)\tilde{T}_{n+p,q}$. This requirement yields $K C_{d-d_0}=d C_d$. 

The deformations discussed above and also those of \eqref{h-barh-H-deformation} involve an index structure which has a shift (by $d, \bar{d}$). In all of these cases, as we explicitly showed, such deformations are trivial ones and can be absorbed in the redefinition of generators. One may show that all such shifts in the indices are trivial deformations. This may be understood  geometrically recalling that the $\mathfrak{bms}_4$ algebra is associated with asymptotic symmetry algebra of 4d flat space and the generators are functions on the 2d celestial sphere \cite{Barnich:2011mi, Barnich:2011ct}. The deformations with the shifted indices are then an inner automorphism of the asymptotic symmetry generating diffeomorphisms; see appendix \ref{appendix-A} for more discussions on this point. Therefore, from now on we only consider deformations with appropriately fixed indices; we do not consider the extra shifts.

\paragraph{With coefficients in \texorpdfstring{$\mathcal{L}$}{L} and \texorpdfstring{$\bar{\mathcal{L}}$}{bar-L}.}

As the \bmst\ case, we can consider deformations of the $[\mathcal{L},T]$ (or $[\mathcal{L},T]$) by $\mathcal{L}$ and $\bar{\mathcal{L}}$  terms:
\begin{equation} 
\begin{split}\label{LT-Ldeform}
 [\mathcal{L}_{m},T_{p,q}] &=(\frac{m+1}{2}-p)T_{p+m,q}+\eta f(m,p)\mathcal{L}_{p+m-1}\delta_{q,0}+\sigma g(m)\bar{\mathcal{L}}_{q-1}\delta_{m+p,0}, \\
 [\bar{\mathcal{L}}_{n},T_{p,q}] &=(\frac{n+1}{2}-p)T_{p,n+q}+\bar{\eta} \bar{f}(n)\mathcal{L}_{p-1}\delta_{n+q,0}+\bar{\sigma} \bar{g}(n,q)\bar{\mathcal{L}}_{n+q-1}\delta_{p,0},
\end{split}\end{equation}
where functions $f,g,\bar{f}$ and $\bar{g}$ are functions to be fixed upon the requirement of closure of the algebra. The index structure of the deformations has been fixed recalling the discussions in last part of the previous subsubsection. 

To find the explicit form of the functions we should consider three different Jacobi identities. The Jacobi $[\mathcal{L}_{m},[\bar{\mathcal{L}}_{n},T_{p,q}]]+{cyclic\ permutation}=0$ leads to one relation for each of $\mathcal{L}$ and $\bar{\mathcal{L}}$ coefficients as
\begin{equation} 
 \delta_{n+q,0}\big( (m-p+1)\bar{f}(n)+(p-\frac{m+1}{2})\bar{f}(n)+(\frac{n+1}{2}-q)f(m,p) \big)=0,\label{barf-f}
\end{equation}
and 
\begin{equation} 
 \delta_{m+p,0}\big( -(n-q+1)g(m)+(\frac{n+1}{2}-q)g(m)+(p-\frac{m+1}{2})\bar{g}(n,q)\big)=0.\label{barg-g}
\end{equation}
From the first relation we have 
\begin{equation} 
(\frac{m+1}{2})\bar{f}(n)=-(\frac{3n+1}{2})f(m,p), \label{barf-f1}
\end{equation}
which suggests that $f(m,n)=a(1+m)$ and $\bar{f}(n)=-a(3n+1)$ and similarly for $\bar{g}(m,n)$ and $g(n)$. 

The next Jacobi we should consider is $[\mathcal{L}_{m},[\mathcal{L}_{n},T_{p,q}]]+{cyclic\ permutation}=0$ which leads to, 
\begin{equation} 
    \delta_{m+p+n,0}\big((\frac{n+1}{2}-p)g(m)+(p-\frac{m+1}{2})g(n)+(n-m)g(m+n)\big) \bar{\mathcal{L}}_{q-1}=0,\label{LT-g}
\end{equation}
and
\begin{equation} 
\begin{split}
   &\delta_{q,0}\big((m-n-p+1)f(n,p)+(\frac{n+1}{2}-p)f(m,p+n)-(n-m-p+1)f(m, p)\\
   &+(p-\frac{m+1}{2})f(n,p+m)+(n-m)f(m+n,p)\big) \mathcal{L}_{m+p-1}=0,\label{LT-f}  
\end{split}
\end{equation}
One may readily verify that $g(n)=-a(1+3n)$ and $f(m,n)=a(1+m)$ respectively solve \eqref{LT-g} and \eqref{LT-f}, as also implied from our previous analysis. 

The last Jacobi we should consider is 
$[T_{p,q},[T_{r,s},\mathcal{L}_{m}]]+{cyclic\ permutation}=0$ which leads to
\begin{equation} 
\begin{split}
   &\delta_{s,0}(\frac{m+r}{2}-p)f(m,r)T_{m+r+p-1,q}+\delta_{r+m,0}(\frac{s}{2}-q)g(m)T_{p,q+s-1}+\\
   &\delta_{q,0}(r-\frac{m+p}{2})f(m,p)T_{m+r+p-1,s}+\delta_{p+m,0}(s-\frac{q}{2})g(m)T_{r,q+s-1}=0.\label{LTT-fg}
\end{split}
\end{equation}
There is a similar equation for $\bar{f}$ and $\bar{g}$ from $[T_{p,q},[T_{r,s},\bar{\mathcal{L}}_{m}]]+{cyclic\ permutation}=0$. The terms with coefficients $g$ should be equal to zero as they are coefficients of different $T_{m,n}$'s. In a similar way the $f(m,n)$ terms should be zero.
So, $[{\cal L}, T]$ cannot be deformed with coefficients in ${\cal L}, \bar{{\cal L}}$.

To summarize this subsection, \bmsf\ algebra can be deformed to a four parameter family of $\mathcal{W}(a,b;\bar{a},\bar{b})$ algebras; where $\mathfrak{bms}_4=\mathcal{W}(\frac12,-\frac12;\frac12,-\frac12)$. $\mathcal{W}(a,b;\bar{a},\bar{b})$ for any value of parameters $a,b;\bar{a},\bar{b}$ 
share a $\mathfrak{witt}\oplus\mathfrak{witt}$ subalgebra spanned by 
${\cal L}_n$ and ${\bar{\cal L}}_n$. In section \ref{sec:5} we will study this family of algebras, its stability and  deformations in more details.


\subsection{ Deformations of commutator of  \texorpdfstring{$[T,T]$}{TT} }\label{ideal-bms}

$[T,T]$ commutator may be deformed in terms involving $T$ or  ${\cal L}$ and  ${\bar{\cal L}}$. In what follows we consider these cases separately. 

\paragraph{With coefficients in \texorpdfstring{$T$}{T}.}
As general case, we can consider the deformation of $[T,T]$ as
\begin{equation} 
 [T_{m,n},T_{p,q}]=G(m,n;p,q)T_{m+p,n+q},\label{TT-T}
\end{equation}
in which $G$ is an  antisymmetric function under the replacements $m\leftrightarrow p$ and $n\leftrightarrow q$. One must check the Jacobi identity $[\mathcal{L}_{r},[T_{m,n},T_{p,q}]]+cyclic\,\,permutations=0$, which yields
\begin{equation}\label{TTT}
    (p-\frac{r+1}{2})G(m,n;p+r,q)+(\frac{r+1}{2}-m)G(p,q;m+r,n)+
    (\frac{r+1}{2}-m-p)G(m,n;p,q)=0.
\end{equation}
For $r=0$, and recalling  $G(m,n;p,q)=-G(p,q;m,n)$, we get
\begin{equation} 
    \big((p-\frac{1}{2})-(\frac{1}{2}-m)+(\frac{1}{2}-m-p)\big)G(m,n;p,q)=0,
\end{equation}
which means that $G(m,n;p,q)=0$. In this way, we have shown that the ideal part of \bmsf\ cannot be deformed by terms with coefficients in $T$, when other commutators are untouched.

\paragraph{With coefficients in \texorpdfstring{$\mathcal{L}$}{L} and \texorpdfstring{$\bar{\mathcal{L}}$}{bar-L}.}\label{L-barL-ideal}
We next consider  deformation of the $[T,T]$ by terms with coefficients in $\mathcal{L}$ and $\bar{\mathcal{L}}$ as 
\begin{equation}\label{A-B--TTL}
 [T_{m,n},T_{p,q}]=A(m,n;p,q)\mathcal{L}_{m+p-1}+B(m,n;p,q)\bar{\mathcal{L}}_{n+q-1}
 \end{equation}
in which the coefficients $A(m,n;p,q)$ and $B(m,n;p,q)$ are antisymmetric under the replacement $m,n\leftrightarrow p,q$.\footnote{Note that the global part of this deformed algebra is always a deformation of 4d Poincar\'e which is  not (necessarily) AdS$_{4}$ algebra $\mathfrak{iso}(3,2)$.} 
The index structure of ${\cal L}$ and ${\bar{\mathcal{L}}}$ in \eqref{A-B--TTL} is chosen recalling discussions of appendix \ref{appendix-A} that the generators may be viewed as fields (operators) on an $S^2$ in Poincar\'e coordinates. Moreover, the Jacobi $[\mathcal{L},[T,T]]+cyclic\,\,permutations=0$ restricts the index of $\mathcal{L}$ to be a linear function of the first indices of $T$. The same argument is obtained for $\bar{\mathcal{L}}$.

One should examine  the Jacobi identities $[\mathcal{L}_{r},[T_{m,n},T_{p,q}]]+cyclic\,\,permutations=0$ and $[T_{r,s},[T_{m,n},T_{p,q}]]+cyclic\,\,permutations=0$. From the first identity for the terms with coefficients $\bar{\mathcal{L}}$  one gets
\begin{equation} 
 \big((p-\frac{r+1}{2})B(m,n;p+r,q)+(\frac{r+1}{2}-m)B(p,q;m+r,n)\big)\bar{\mathcal{L}}_{n+q-1}=0.\label{B-eq0}
 \end{equation}
For $r=0$ and recalling the antisymmetry of $B$ function, we find $(m+p-1) B(m,n;p,q)=0$ and therefore, $B(m,n;p,q)=B_0(n,q) \delta_{m+p,1},$ where $B_0(n,q)=-B_0(q,n)$. Next we plug this form of $B$ back into \eqref{B-eq0} to obtain
$-2r B_0(n,q)\delta_{m+p+r,1}=0$ which implies $B_0(n,q)=0$ and hence $B$ should vanish.
A similar argument works for $A(m,n;p,q)$ when we consider the Jacobi $[\mathcal{L}_{r},[T_{m,n},T_{p,q}]]+cyclic\,\,permutations=0$ and hence $A=B=0$.

To summarize this section, we have shown that $\mathfrak{bms}_4$ algebra admits non-trivial infinitesimal deformation only in $[{\cal L}, T]$ and $[\bar{\cal L}, T]$ parts of the algebra by coefficients in $T$. Therefore, the only allowed infinitesimal deformations of the $\mathfrak{bms}_4$ algebra is ${\cal W}(a,b;\bar{a},\bar{b})$.

\section{Most general formal deformations of \texorpdfstring{$\mathfrak{bms}_4$}{BMS4}  algebra}\label{sec:4}

Here we complete the analysis of previous section by showing that (1) the infinitesimal deformations of \bmsf\ into ${\cal W}(a,b;\bar{a},\bar{b})$ are also formal deformations and (2) there are no other deformations possible when we consider simultaneous deformations of two or more commutators. To this end, let us consider the schematic form of the most general deformations of  $\mathfrak{bms}_4$ in which all deformations are turned on simultaneously 
\begin{equation} 
\begin{split}
 & [\mathcal{L},\mathcal{L}]=\mathcal{L}+hT, \\
 & [\bar{\mathcal{L}},\bar{\mathcal{L}}]=\bar{\mathcal{L}}+\bar{h}T,\\
 &[{\mathcal{L}},\bar{\mathcal{L}}]=HT,\\
 &[\bar{\mathcal{L}},T]=T+\bar{K}T+\bar{f}\mathcal{L}+\bar{g}\bar{\mathcal{L}},\\
 &[\mathcal{L},T]=T+KT+f\mathcal{L}+g\bar{\mathcal{L}},\\
 &[T,T]=GT+A\bar{\mathcal{L}}+B\mathcal{L},
\end{split}\label{most-deform}
\end{equation}

The Jacobi $[\mathcal{L},[\mathcal{L},\mathcal{L}]]+cyclic\,\,permutations=0$ (and $[\bar{\mathcal{L}},[\bar{\mathcal{L}},\bar{\mathcal{L}}]]+cyclic\,\,permutations=0$) leads to some relations just for $h$ (and $\bar{h}$), in accord with the analysis of section \ref{Witt-deformations-sec},  have solution $h,\bar{h}=constant$ up to first order in deformation parameters (infinitesimal deformation). 
The Jacobi $[\mathcal{L},[\mathcal{L},\bar{\mathcal{L}}]]+cyclic\,\,permutations=0$ (and $[\bar{\mathcal{L}},[\bar{\mathcal{L}},\mathcal{L}]]+cyclic\,\,permutations=0$) up to first order just leads to \eqref{H-h} and \eqref{H-barh}; deformations in the $[{\cal L}, T], [\bar{\cal L}, T], [T, T]$ parts do not alter the equations on $h,\bar{h}$ and $H$. Therefore, there are no non-trivial deformations coming from this sector.

The Jacobi $[\mathcal{L},[\bar{\mathcal{L}},T]]+cyclic\,\,permutations=0$ up to first order just leads to the constraints \eqref{barf-f} and \eqref{barg-g}. 
  
The Jacobi $[\mathcal{L},[\mathcal{L},T]]+cyclic\,\,permutations=0$ (and $[\bar{\mathcal{L}},[\bar{\mathcal{L}},T]]+cyclic\,\,permutations=0$) up to first order just leads to the constraints \eqref{eq-K}, \eqref{LT-g} and \eqref{LT-f}.

The Jacobi $[T,[T,T]]+cyclic\,\,permutations=0$ does not lead to any constraints up to first order in the deformation parameter.

Finally, the Jacobi $[T,[T,\mathcal{L}]]+cyclic\,\,permutations=0$ (and $[T,[T,\bar{\mathcal{L}}]]+cyclic\,\,permutations=0$) up to first order leads to the following  three independent relations
\begin{equation} 
\begin{split}
  &\delta_{s,0}(\frac{m+r}{2}-p)f(m,r)T_{m+r+p-1,q}+\delta_{r+m,0}(\frac{s}{2}-q)g(m)T_{p,q+s-1}+\\
   &\delta_{q,0}(r-\frac{m+p}{2})f(m,p)T_{m+r+p-1,s}+\delta_{p+m,0}(s-\frac{q}{2})g(m)T_{r,q+s-1}+ \\
  &\big((r-\frac{m+1}{2})G(p,q;r+m,s)+(\frac{m+1}{2}-p)G(r,s;p+m,q)+\\
    &(\frac{m+1}{2}-(r+p+d))G(p,q;r,s)\big)T_{m+p+r,s+q}=0,
\end{split}\label{most-deform-LTT1}
\end{equation}
\begin{equation} 
\begin{split}
 & \big[(r-\frac{m+1}{2})A(p,q;m+r,s)+(\frac{m+1}{2}-p)A(r,s;p+m,q)+\\
  +& (m-p-r+1)A(p,q;r,s)\big]\mathcal{L}_{m+p+r-1}=0, 
\end{split}\label{most-deform-LTT2}
\end{equation}
and 
 \begin{equation} \label{most-deform-LTT3}
  \big((r-\frac{m+1}{2})B(p,q;m+r,s)+(\frac{m+1}{2}-p)B(r,s;p+m,q)\big)\bar{\mathcal{L}}_{s+q-1}=0. 
\end{equation}
As discussed, \eqref{most-deform-LTT3} leads to $B(p,q;r,s)=0$. A similar argument (analyzing \eqref{most-deform-LTT2} for $m=0$) yields $A(p,q;r,s)=0$. So we should just analyze \eqref{most-deform-LTT1}. Since $T_{m,n}$ for different $m,n$ are linearly independent, a careful analysis of the indices of $T$ generators in \eqref{most-deform-LTT1} reveals that  $f(m,n),g(m)$ and $G(m,n;p,q)$ should all vanish. To summarize, turning on deformations simultaneously, up to the first order, does not yield any new deformation other than ${\cal W}(a,b;\bar{a},\bar{b})$ algebra. 

\paragraph{Integrability, obstructions and formal deformation.}\label{Integrability-Wab}

We have shown in the previous section that the most general infinitesimal nontrivial deformation of \bmsf\ is $\mathcal{W}(a,b;\bar{a},\bar{b})$. Now, we would like to explore integrability of these deformations and check if they are formal deformations. As in the case of \bmst\ discussed in \cite{Parsa:2018kys} one only needs to consider the relation
  \begin{multline}
(\frac{n+1}{2} - p) K(m, p + n) + ( \frac{m+1}{2}-p-n) K(n, p) + K(n,p)K(m,n+p)+ (p-\frac{m+1}{2}) K(
   n, p+m) +\\  +
  (p+m - \frac{n+1}{2}) K(m,p)-K(m,p)K(n,p+m)+(n-m) K(m+n,p)=0.\label{formal-W}
\end{multline} 
which is satisfied by $K(m,n)=\alpha+\beta m$.\footnote{We note that the most general solution of \eqref{formal-W} is $\alpha+\beta m$ plus the solution given in \eqref{K-redef}. However, the latter is trivial deformation and may be absorbed in redefinition of $T$ as in \eqref{T-redef}. A similar pattern was also found in the 3d case, \emph{cf.} section 4.2 of \cite{Parsa:2018kys}.} (Considering the Jacobi $[\bar{{\cal L}}_{m},[{\cal L}_{n},T_{p,q}]]+{cyclic\ permutations}=0$ does not change this result.) This means that the obtained infinitesimal deformation is integrable and can be extended to formal deformation.

{
Analysis of previous and this section may be summarized in the  the following  theorem: 
\paragraph{Theorem 4.1} {\it The most general formal deformation of \bmsf\ is $\mathcal{W}_{4}(a,b;\bar{a},\bar{b})$ algebra}. }
\section{On \texorpdfstring{$\mathcal{W}(a,b;\bar{a},\bar{b})$}{W4}  algebra, its subalgebras and  deformations}\label{sec:5}

We have introduced $\mathcal{W}(a,b;\bar{a},\bar{b})$ which appears as formal deformation of \bmsf\ and here we would like to study this algebra a bit more. We first analyze its global subalgebras and then consider possible deformations of $\mathcal{W}(a,b;\bar{a},\bar{b})$, particularly for special values of $a,b,\bar{a},\bar{b}$ parameters. Before starting we note that, as in the \bmst\ and $W(a,b)$ algebra cases \cite{Parsa:2018kys}, 
\begin{itemize}
\item ${\cal W}(a,b;\bar{a}, \bar{b})$ and  ${\cal W}(\bar{a}, \bar{b}; a,b)$ algebras are isomorphic. 
\item the range of $a, \bar{a}$ parameters may be limited to $[-1/2,1/2)$, as   $a=k+r, k\in\mathbb{Z}$  and $a=r$ cases can be related  by just a shift in the index of the associated $T_{m,n}$ generator, $T_{m,n}\to T_{m-k,n}$, and simiarly for the $\bar{a}$. 
\item ${\cal W}(a,b;\bar{a},\bar{b})$ and ${\cal W}(-a,b;\bar{a},\bar{b})$ algebras are isomorphic, as renaming ${\cal L}_m\to -{\cal L}_{-m}$ and $T_{p,q}\to T_{-p,q}$ relates these two algebras. Therefore, one may restrict the range of $a$ and $\bar{a}$ parameters to $[0,1/2]$. 
\end{itemize}

\subsection{Subalgebras of \texorpdfstring{$\mathcal{W}(a,b;\bar{a},\bar{b})$}{W4}}
 Irrespective of the values of $a,b;\bar{a},\bar{b}$ parameters, all $\mathcal{W}(a,b;\bar{a},\bar{b})$ algerbas share a $\mathfrak{witt}\oplus\mathfrak{witt}$ subalgebra spanned by 
${\cal L}_n$ and $\bar{\mathcal{L}}_n$. This subalgebra in turn has a Lorentz subalgebra $\mathfrak{so}(3,1)=\mathfrak{sl}(2\mathbb{R})_L\oplus \mathfrak{sl}(2\mathbb{R})_R$ spanned by $\mathcal{L}_{0},\mathcal{L}_{\pm1},\ \bar{\mathcal{L}}_{0}, \bar{\mathcal{L}}_{\pm 1}$. Depending on the values of the four parameters, some $T_{m,n}$ generators may also be a part of this global subalgebra, e.g. as discussed in the previous section for $\mathfrak{bms}_4=\mathcal{W}(-\frac12,-\frac12;-\frac12,-\frac12)$,  $T_{0,0}, T_{0,1}, T_{1,0}, T_{1,1}$ are the other four generators which turn the global subalgebra to 4d Poincar\'e $\mathfrak{iso}(3,1)$. To verify which of $T_{m,n}$ appear in the global subalgebra, we consider the relevant commutators:
\begin{equation} 
\begin{split}
 &[\mathcal{L}_{+1},T_{p,q}]=-(p+b+a)T_{p+1,q}, \\
 & [\mathcal{L}_{0},T_{p,q}]=-(p+a)T_{p,q},\\
 &[\mathcal{L}_{-1},T_{p,q}]=-(p-b+a)T_{p-1,q},
\end{split}\label{L-Tglobal}
\end{equation}
and 
\begin{equation} 
\begin{split}
 & [\bar{\mathcal{L}}_{+1},T_{p,q}]=-(q+\bar{b}+\bar{a})T_{p,q+1}, \\
 & [\bar{\mathcal{L}}_{0},T_{p,q}]=-(q+\bar{a})T_{p,q},\\
 &[\bar{\mathcal{L}}_{-1},T_{p,q}]=-(q-\bar{b}+\bar{a})T_{p,q-1}.
\end{split}\label{barL-Tglobal}
\end{equation}
The above commutations close off for finite number of $T_{m,n}$ generators only for three cases, $a=b=0$, $a=0, b=-1$ and $a=b=-1/2$, and similarly for $\bar{a},\bar{b}$. For generic values of $a,b$ other than these two cases the global part (finite subalgebra) of ${\cal W}(a,b;\bar{a},\bar{b})$ is just $\mathfrak{so}(3,1)$. We have therefore, nine independent special cases for which the global part of the algebra is bigger than the Lorentz algebra $\mathfrak{so}(3,1)$:
\begin{enumerate}
\item ${\cal W}(-1/2,-1/2;-1/2,-1/2)$ which is nothing but the \bmsf\ and its global part is 4d Poincar\'e algebra $\mathfrak{iso}(3,1)$.
\item ${\cal W}(0,0;0,0)$, where $T_{0,0}$ falls into the global part and hence we have $\mathfrak{so}(3,1)\oplus \mathfrak{u}(1)$ global algebra.
\item ${\cal W}(0,-1;0,0)$ (or ${\cal W}(0,0;0,-1)$), where $T_{r,0},\ r=0,\pm 1$ (or $T_{0,r}$) fall into the global part and the global subalgebra is $\mathfrak{iso}(2,1)\oplus \mathfrak{sl}(2,\mathbb{R})_R$ (or $\mathfrak{iso}(2,1)\oplus \mathfrak{sl}(2,\mathbb{R})_L$). Generators of this algebra may be represented in usual Minkowski tensors: ${J}_{\mu\nu}, F^+_{\mu\nu}$, where $F^+_{\mu\nu}$ is a self-dual (anti-self-dual) anti-symmetric object, representing the $T_{r,0}$:
\be\begin{split}
[J_{\mu\nu}, J_{\alpha\beta}] &= i(\eta_{\mu \alpha} J_{\nu \beta}+\eta_{\beta \mu}J_{\alpha \nu}-\eta_{\nu \alpha}J_{\mu \beta}-\eta_{\beta \nu}J_{\alpha \mu}),\\
[{J}_{\mu\nu}, F^+_{\alpha\beta}] &={i(\eta_{\mu \alpha} F^+_{\nu \beta}+\eta_{\beta \mu}F^+_{\alpha\nu}-\eta_{\nu \alpha}F^+_{\mu \beta}-\eta_{\beta \nu}F^+_{\alpha\mu}),} \\
[F^+_{\mu\nu},F^+_{\alpha\beta}] &= 0. 
\end{split}
\ee
\item ${\cal W}(0,-1;0,-1)$ where the global subalgebra involves nine generators $T_{r\bar{s}}, r,\bar{s}=0,\pm 1$. This 15 dimensional algebra which generated by ${\cal L}_r, {\bar{\mathcal{L}}}_{\bar s}, T_{r\bar{s}}$, $r,\bar{s}=0,\pm 1$. These generators may be gathered in a traceless Lorentz two-tensor $K_{\mu\nu}$, where its antisymmetric part is Lorentz generators $J_{\mu\nu}$ and its symmetric part captures $T_{r\bar{s}}$ and satisfy the algebra
\begin{equation}
   i [K_{\mu\nu}, K_{\alpha\beta}]=\eta_{\mu\beta}K_{\alpha\nu}-\eta_{\alpha\nu}K_{\mu\beta}.
\end{equation}
The above algebra is $\mathfrak{so}(3,1)\inplus_{ad} \mathfrak{T}$ subalgebra where $\mathfrak{T}$ denotes the ideal part which is spanned by $T_{r\bar{s}}$; the $T_{r\bar{s}}$ are in the bi-adjoint of the $\mathfrak{sl}(2,\mathbb{R})_L\oplus \mathfrak{sl}(2,\mathbb{R})_R$.

\item ${\cal W}(0,0;-1/2,-1/2)$ (or ${\cal W}(-1/2,-1/2;0,0)$), where $T_{0,\alpha}, \alpha=0,1$ (or $T_{\alpha,0}$) are also in the global algebra which is eight dimensional. This global algebra is  $\mathfrak{sl}(2,\mathbb{R})\oplus \mathfrak{sch}_2$ algebra where $\mathfrak{sch}_2$ denotes the $2d$ Schr\"{o}dinger algebra without central element.

\item ${\cal W}(0,-1;-1/2,-1/2)$ (or ${\cal W}(-1/2,-1/2;0,-1)$) where its global subalgebra is $\mathfrak{so}(3,1) \inplus_{ad} \mathfrak{T}_{\alpha r}$, where $\mathfrak{T}_{\alpha r}$ is  spanned by $T_{\alpha r}, \alpha=0,1, \ r=0,\pm 1$
which are in the vector and spinor representation of $\mathfrak{sl}(2,\mathbb{R})_L\oplus \mathfrak{sl}(2,\mathbb{R})_R$ subalgebra.

\end{enumerate}

\paragraph{Some other infinite dimensional subalgebras of $\mathcal{W}(a,b;\bar{a},\bar{b})$.} Besides the above examples, one may consider other infinite dimensional subalgebras of $\mathcal{W}(a,b;\bar{a},\bar{b})$. The simplest of these subalgebras are $\mathfrak{witt}\oplus\mathfrak{witt}$ or $W(a,b)_p$ algebras generated by ${\cal L}_n, T_{m, p}$ where $p$ is a fixed (but arbitrary) number.  As a subalgebra of ${\cal W}(a,b; 0, -1)$ one may consider the one generated by ${\cal L}_n, T_{m, r},\bar{\mathcal{L}}_r\  (r=0,\pm 1)$. In special case $a=b=0$ one may view this as a $\mathfrak{u}(1)^3$ Kac-Moody algebra where the three currents fall into triplet representation  of the $\mathfrak{sl}(2,\mathbb{R})_R$ generated by  ${\bar{\mathcal{ L}}}_r$. This latter may be viewed as an ``internal symmetry'' of the Kac-Moody part. 

\subsection{Deformations of generic \texorpdfstring{$\mathcal{W}(a,b;\bar{a},\bar{b})$}{W4}  algebra}

As discussed formal deformations of \bmsf\ algebra yields the four parameter family $\mathcal{W}(a,b;\bar{a},\bar{b})$ algebra. As such, one expects this algebra to be rigid. However, a more careful look into the analysis of previous section also reveals that deformations may move us in the $a,b;\bar{a},\bar{b}$ plane. This is very similar to the 3d example of $W(a,b)$ discussed in \cite{Parsa:2018kys}, and is what we will explore more explicitly in this section. Here we use $\mathcal{W}(a,b;\bar{a},\bar{b})$ in two different meanings, which will hopefully be clear from the context: (1) $\mathcal{W}(a,b;\bar{a},\bar{b})$ for a given (but generic) value of the four parameter. This latter denotes a specific algebra; (2) $\mathcal{W}(a,b;\bar{a},\bar{b})$ as ``family'' of algebras for different values of the parameters. 

While the $\mathcal{W}(a,b;\bar{a},\bar{b})$ family is expected it to be rigid, as in the 3d example of $W(a,b)$ discussed in \cite{Parsa:2018kys}, there could be special values of parameters where one can deform the algebra to other families of algebras. These special points, as we will see, correspond to cases with larger global part discussed in the previous subsection.
To make a formal analysis of $\mathcal{W}(a,b;\bar{a},\bar{b})$ as in the previous section we consider all possible deformations of its commutators and check the closure conditions.

\paragraph{Deformations of $\mathfrak{witt}\oplus \mathfrak{witt}$ part.}\label{witt-witt-T} One can deform this sector of  $\mathcal{W}(a,b;\bar{a},\bar{b})$ algebra as
\begin{equation} 
\begin{split}
     &[\mathcal{L}_{m},\mathcal{L}_{n}]=(m-n)\mathcal{L}_{m+n}+(m-n)\sum_{d,\bar{d}} h^{d,\bar{d}}(m,n)T_{m+n+d,\bar{d}},\\
    &[\bar{\mathcal{L}}_{m},\bar{\mathcal{L}}_{n}]=(m-n)\bar{\mathcal{L}}_{m+n}+(m-n)\sum_{d,\bar{d}} \bar{h}^{d,\bar{d}}(m,n)T_{d,m+n+\bar{d}},\\
    &[\mathcal{L}_{m},\bar{\mathcal{L}}_{n}]=\sum_{d,\bar{d}} H^{d,\bar{d}}(m,n)T_{m+d,n+\bar{d}},\label{w-w-W4-deform}
\end{split}
\end{equation}
in which $h^{d,\bar{d}}(m,n),\bar{h}^{d,\bar{d}}(m,n)$ are symmetric and $H^{d,\bar{d}}(m,n)$ an arbitrary functions. 
As the first step, one considers the Jacobi $[\mathcal{L}_{m},[\mathcal{L}_{n},\mathcal{L}_{l}]]+cyclic\,\,permutations=0$ which leads to 
\begin{equation} 
\begin{split}
    &\sum_{d,\bar{d}}\{\big((n-l)(m-n-l)h^{d,\bar{d}}(m,n+l)+(n-l)(-bm-a-n-l-d)h^{d,\bar{d}}(n,l)+\\
    &+(l-m)(n-m-l)h^{d,\bar{d}}(n,m+l)+(l-m)(-bn-a-m-l-d)h^{d,\bar{d}}(l,m)+\\
    &+(m-n)(l-m-n)h^{d,\bar{d}}(l,m+n)+(m-n)(-bl-a-m-n-d)h^{d,\bar{d}}(m,n)\big)T_{m+n+l+d,\bar{d}}\}=0,\label{h-a,b}
\end{split}
\end{equation}
and the same relation for $\bar{h}^{d,\bar{d}}(m,n)$. The only solution to the above equations for generic  $a,b$ is $h^{d,\bar{d}}(m,n)=constant$ (and $\bar{h}^{d,\bar{d}}(m,n)=constant$).

The next two Jacobi identities to analyze are $[\mathcal{L}_{m},[\mathcal{L}_{n},\bar{\mathcal{L}}_{l}]]+cyclic\,\,permutations=0$ and 
$[\bar{\mathcal{L}}_{m},[\bar{\mathcal{L}}_{n},\mathcal{L}_{l}]]+cyclic\,\,permutations=0$, which yield
\begin{align}
     &\bigg(-(a+bm+n+d)H^{d,\bar{d}}(n,l)+(a+bn+m+d)H^{d,\bar{d}}(m,l)+(n-m)H^{d,\bar{d}}(m+n,l)+\cr
     &(n-m)(\bar{a}+\bar{b}l+\bar{d})h^{d,\bar{d}}(m,n)\bigg)T_{m+n+d,l+\bar{d}}=0,\label{H-h-ab}\\ \;\;\;\cr
   &\bigg((\bar{a}+\bar{b}m+n+\bar{d})H^{d,\bar{d}}(l,n)-(\bar{a}+\bar{b}n+m+\bar{d})H^{d,\bar{d}}(l,m)+(m-n)H^{d,\bar{d}}(l,m+n)+\cr
   &(n-m)(a+bl+d)\bar{h}^{d,\bar{d}}(m,n)\bigg)T_{l+d,m+n+\bar{d}}=0.\label{H-barh-ab}  
\end{align}
As in the \bmsf\  case one can obtain the most general form of the $H^{d,\bar{d}}(m,n)$ as
\begin{equation}
    H^{d,\bar{d}}(m,n)=H^{d,\bar{d}}_0 (bm+a+d)(\bar{b}n+\bar{a}+\bar{d})+\frac{\bar{h}}{\bar{b}}(bm+a+d)-\frac{h}{b}(\bar{b}n+\bar{a}+\bar{d}).\label{general-H-w4}
\end{equation}

One can then show that for generic $a,b,\bar{a},\bar{b}$ through the redefinitions \eqref{Redef-XY} in which $X^{d,\bar{d}}(m)$, $Y^{d,\bar{d}}(m)$ are changed to
\begin{equation}
    X^{d,\bar{d}}(m)=H_0^{d,\bar{d}}(bm+a+d)+\frac{h^{d,\bar{d}}}{b},\,\,\, Y^{d,\bar{d}}(m)=-H_0^{d,\bar{d}}(\bar{b}m+\bar{a}+\bar{d})+\frac{\bar{h}^{d,\bar{d}}}{\bar{b}},\label{Redef-XY-w4}
\end{equation}
these deformations can be reabsorbed and hence they are all trivial. However, as we see for the special case of $b=0$ or $\bar{b}=0$ these redefinitions are not well-defined and hence there remains a non-trivial deformation for  $\mathcal{W}(a,b;0,0)$, $\mathcal{W}(0,0;\bar{a},\bar{b})$ and  $\mathcal{W}(0,0;0,0)$ cases which we discuss below separately.

 \paragraph{Deformations of \texorpdfstring{$[\mathcal{L},T]$}{LT-W} and \texorpdfstring{$[\bar{\mathcal{L}},T]$}{barLT-W} commutators.}\label{K-f-g-w4}
The most general deformations in this sector of  $\mathcal{W}(a,b;\bar{a},\bar{b})$ algebra is 
\begin{equation} 
\begin{split}
    &[\mathcal{L}_{m},T_{p,q}]=-(a+bm+p)T_{p+m,q}+K(m,p)T_{p+m,q}++\eta f(m,p)\mathcal{L}_{p+m-1}\delta_{q,0}+\sigma g(m)\bar{\mathcal{L}}_{q-1}\delta_{m+p,0},\\
 &[\bar{\mathcal{L}}_{n},T_{p,q}]=-(\bar{a}+\bar{b}n+q)T_{p,q+n}+\bar{K}(n,q)T_{p,n+q}+\bar{\eta} \bar{f}(n)\mathcal{L}_{p-1}\delta_{n+q,0}+\bar{\sigma} \bar{g}(n,q)\bar{\mathcal{L}}_{n+q-1}\delta_{p,0},\label{L-barL-T-W4-deform}
\end{split}
\end{equation}
in which $K,\bar{K},f,g,\bar{f}$ and $\bar{g}$ are arbitrary functions. 
As the first step, we considers the Jacobi $[\mathcal{L}_{m},[\mathcal{L}_{n},T_{p,q}]]+cyclic\,\,permutations=0$, leading to 
\begin{multline}\label{K-w4}
(-a-bn - p) K(m, p + n) + ( -a-bm-p-n) K(n, p) + (p+a+bm) K(
   n, p+m) +\\  +
  (p+m+a+bn) K(m,p)+ (n-m) K(m+n,p)=0.
\end{multline}
and the same relation for $\bar{K}(m,n)$. One can solve the latter to get $K(m,n)=\alpha+\beta m$ and similar result for $\bar{K}(m,n)$. From previous Jacobi one also obtains two other relations for $f(m,n)$ and $g(m,n)$ as 
\begin{equation} 
    \delta_{m+p+n,0}\big((-a-bn-p)g(m)+(p+a+bm)g(n)+(n-m)g(m+n)\big) \bar{\mathcal{L}}_{q-1}=0,\label{LT-g-w4}
\end{equation}
and
\begin{equation} 
\begin{split}
   &\delta_{q,0}\big((m-n-p+1)f(n,p)+(-a-bn-p)f(m,p+n)-(n-m-p+1)f(m, p)\\
   &+(p+a+bm)f(n,p+m)+(n-m)f(m+n,p)\big) \mathcal{L}_{m+p-1}=0,\label{LT-f-w4}  
\end{split}
\end{equation}
and similar relations for $\bar{f}(m,n)$ and $\bar{g}(m,n)$ can be obtained from the Jacobi $[\bar{\mathcal{L}}_{m},[\bar{\mathcal{L}}_{n},T_{p,q}]]+cyclic\,\,permutations=0$.

On the other hand, the Jacobi $[\mathcal{L}_{m},[\bar{\mathcal{L}}_{n},T_{p,q}]]+cyclic\,\,permutations=0$ leads to one relation for each of $\mathcal{L}$ and $\bar{\mathcal{L}}$ coefficients as
\begin{equation} 
 \delta_{n+q,0}\big( (m-p+1)\bar{f}(n)+(p+a+bm)\bar{f}(n)+(-a-bn-q)f(m,p) \big)=0,\label{barf-f-w4}
\end{equation}
and 
\begin{equation} 
 \delta_{m+p,0}\big( -(n-q+1)g(m)+(-a-bn-q)g(m)+(p+a+bm)\bar{g}(n,q)\big)=0.\label{barg-g-w4}
\end{equation}
From the first relation we have 
\begin{equation} 
(m(b+1)+a+1)\bar{f}(n)=(a+n(b-1))f(m,p), \label{barf-f1-w4}
\end{equation}
which suggests that $f(m,n)=c(m(b+1)+a+1)$ and $\bar{f}(n)=-c(a+n(b-1))$ where $c=constant$ and similarly for $\bar{g}(m,n)$ and $g(n)$ which is in agreement with \eqref{barf-f1-w4} obtained for $a=b=\frac{-1}{2}$ case.

The last Jacobi we should consider is 
$[T_{p,q},[T_{r,s},\mathcal{L}_{m}]]+\text{cyclic permutation}=0$ which leads to
\begin{equation} 
\begin{split}
   &\delta_{s,0}(a+b(m+r-1)+p)f(m,r)T_{m+r+p-1,q}+\delta_{r+m,0}(\bar{a}+\bar{b}(s-1)+q)g(m)T_{p,q+s-1}-\\
   &\delta_{q,0}(a+b(m+p-1)+r)f(m,p)T_{m+r+p-1,s}-\delta_{p+m,0}(\bar{a}+\bar{b}(q-1)+s)g(m)T_{r,q+s-1}=0.\label{LTT-fg-w4}
\end{split}
\end{equation}
There is a similar equation for $\bar{f}$ and $\bar{g}$ from $[T_{p,q},[T_{r,s},\bar{\mathcal{L}}_{m}]]+\text{cyclic permutation}=0$. The terms with coefficients $g$ should be equal to zero as they are coefficients of different $T_{m,n}$'s. In a similar way $f(m,n)$ terms should be zero.

\paragraph{Deformations of  \texorpdfstring{$\mathcal{W}(a,b;\bar{a},\bar{b})$}{W4}'s ideal part: }
\begin{equation} 
 i[T_{m,n},T_{p,q}]=G(m,n;p,q)T_{m+p,n+q}+A(m,n;p,q)\mathcal{L}_{m+p-1}+B(m,n;p,q)\bar{\mathcal{L}}_{n+q-1},\label{TT-L}
\end{equation}
in which $G,A$ and $B$ are  antisymmetric functions under  $m\leftrightarrow p$ and $n\leftrightarrow q$. The Jacobi identities $[\mathcal{L}_{r},[T_{m,n},T_{p,q}]]+cyclic\,\,permutations=0$ and  $[\bar{\mathcal{L}}_{r},[T_{m,n},T_{p,q}]]+cyclic\,\,permutations=0$ yield two relations for $G$,
\begin{equation}\label{TTT1-w4}
    (p+a+br)G(m,n;p+r,q)-(a+br+m)G(p,q;m+r,n)-
    (a+br+m+p)G(m,n;p,q)=0,
\end{equation}
and
 \begin{equation} 
   (\bar{a}+\bar{b}r+q)G(m,n;p,q+r)-
    (\bar{a}+\bar{b}r+n)G(p,q;m,n+r)-(\bar{a}+\bar{b}r+n+q)G(m,n;p,q)=0,\label{TTT2-w4}
 \end{equation}
and two relations for $A$,
\begin{equation} 
 \big((r-m-p+1)A(m,n;p,q)+(p+a+br)A(m,n;p+r,q)+(m+a+br)A(m+r,n;p,q)\big)\mathcal{L}_{m+p+r-1}=0, \label{A-eq0-w4}
 \end{equation}
and 
\begin{equation} 
 \big((q+\bar{a}+\bar{b}r)A(m,n;p,q+r)+(-\bar{a}-\bar{b}r-n)A(p,q;m,n+r)\big)\mathcal{L}_{m+p-1}=0.\label{A-eq1-w4}
 \end{equation}
The same relation can be obtained from these two Jacobi for $B(m,n;p,q)$.

One can check that our argument in section \ref{ideal-bms} also works for deformations of $\mathcal{W}(a,b;\bar{a},\bar{b})$'s in this sector, in the sense that for generic $a,b,\bar{a}$ and $\bar{b}$ it does not admit any non-trivial deformation in its ideal part. 

\paragraph{The most general deformations of \texorpdfstring{$\mathcal{W}(a,b;\bar{a},\bar{b})$}{W4} algebra.}
When we consider the most general deformations of $\mathcal{W}(a,b;\bar{a},\bar{b})$ algebra simultaneously as in the \bmsf\ case of \eqref{most-deform}, one can verify that except one case, similar to \bmsf\ case, all Jacobi identities lead to the relations obtained in the above. Similarly the \bmsf\ case, the only case we must study is the Jacobi $[T,[T,\mathcal{L}]]+cyclic\,\,permutations=0$ and $[T,[T,\bar{\mathcal{L}}]]+cyclic\,\,permutations=0$ which leads to the relations \eqref{A-eq0-w4}, \eqref{A-eq1-w4} and their analogue for $B$ and sum of two relations \eqref{TTT1-w4} and \eqref{LTT-fg-w4}. One can then show that these two latter are independently equal to zero. In this way the most general deformations of $\mathcal{W}(a,b;\bar{a},\bar{b})$ are restricted to that induces by $K(m,n)$ and takes the $\mathcal{W}(a,b;\bar{a},\bar{b})$ to $\mathcal{W}({a'},{b'};{\bar{a}'},{\bar{b}'})$ with shifted parameters.  

{The following theorem summarizes our above discussion as: 
\paragraph{Theorem 5.1} {\it The family of $\mathcal{W}_{4}(a,b;\bar{a},\bar{b})$ algebra for generic values of the four parameters is stable (rigid) algebra}. }

\subsection{Deformations of special \texorpdfstring{$\mathcal{W}$}{W} algebras}

As discussed while in generic points of the parameter space of ${\cal W}$ algebra they are rigid, there are special points in the parameter space where the algebra is not rigid and may admit other deformations. In this subsection we discuss such special cases.

\paragraph{Deformations of \texorpdfstring{$\mathcal{W}(a,b;0,0)$}{u(1)Kac-Moody}.}
As discussed for $a=b=0$ case the $h$-deformations (\emph{cf}. \eqref{w-w-W4-deform}) become non-trivial and cannot be absorbed into a redefinition of generators. One can go through the Jacobi's of the previous subsection and verify allowed deformations. Here we do not repeat the analysis and just present the final result. The most general \emph{new} non-trivial deformation of  $\mathcal{W}(a,b;0,0)$ is:
\begin{equation} 
\begin{split}
    &[\mathcal{L}_{m},\mathcal{L}_{n}]=(m-n)\mathcal{L}_{m+n},\\
    &[\bar{\mathcal{L}}_{m},\bar{\mathcal{L}}_{n}]=(m-n)\bar{\mathcal{L}}_{m+n}+\bar{\nu} \ (m-n) T_{0,m+n},\\
    &[\mathcal{L}_{m},\bar{\mathcal{L}}_{n}]=0,\\
    &[\mathcal{L}_{m},T_{p,q}]=-(p+bm+a)T_{p+m,q},\\
 &[\bar{\mathcal{L}}_{n},T_{p,q}]=-(q)T_{p,q+n},\\
 &[T_{m,n},T_{p,q}]=0.\label{u(1)-u(1)}
\end{split}
\end{equation}
Some comments are in order:
\begin{itemize}
    \item Among the ${\bar h}^{d,\bar{d}}$ deformations only $\bar{d}=0$ terms remain. The others are still trivial or do not satisfy Jacobi identity.
    \item The possibility of moving in the $(a,b, \bar{a},\bar{b})$ parameter space via the deformation still exists and the above is the new non-trivial deformation which did not exist for $\bar{a},\bar{b}\neq 0$ point. That is, the $\bar{\nu}$ deformation and moving in $\bar{a},\bar{b}$ plane are not mutually inclusive.
\item   $[T_{m,n},T_{p,q}]$ cannot be deformed. Once again, $[\mathcal{L}_{m},T_{p,q}],\ [\bar{\mathcal{L}}_{n},T_{p,q}]$ can only be deformed into 
those of $\mathcal{W}(a,b;\bar{a},\bar{b})$ algebra, but in that case we need to set $\nu=0$.

\item Here we are considering $a,b\neq 0$ case. The special case of $\mathcal{W}(0,0;0,0)$ will be discussed next.

\end{itemize}

\paragraph{The special case of $\mathcal{W}(0,0;0,0)$.} In this case one can deform the algebra into a generic point in the  $(a,b, \bar{a},\bar{b})$ parameter space or alternatively deform the $[\mathcal{L}_{m},\mathcal{L}_{n}]$ (or $[\bar{\mathcal{L}}_{m},\bar{\mathcal{L}}_{n}]$) with the coefficients in $T_{m+n,0}$ (or in $T_{0,m+n})$ while moving in $\bar{a},\bar{b}$ (or $a,b$) plane; or turn on the two $[\mathcal{L}_{m},\mathcal{L}_{n}]$ and $[\bar{\mathcal{L}}_{m},\bar{\mathcal{L}}_{n}]$ deformations without moving in the parameter space. These possibilities are mutually exclusive and the last deformation has the explicit form:
\begin{equation} 
\begin{split}
    &[\mathcal{L}_{m},\mathcal{L}_{n}]=(m-n)\mathcal{L}_{m+n}+\nu\ (m-n)T_{m+n,0},\\
    &[\bar{\mathcal{L}}_{m},\bar{\mathcal{L}}_{n}]=(m-n)\bar{\mathcal{L}}_{m+n}+\bar{\nu}\ (m-n)T_{0,m+n},\\
    &[\mathcal{L}_{m},\bar{\mathcal{L}}_{n}]=H_{0}(\alpha+\beta m)(\bar{\alpha}+\bar{\beta}n),\\
    &[\mathcal{L}_{m},T_{p,q}]=(-p)T_{p+m,q},\\
 &[\bar{\mathcal{L}}_{n},T_{p,q}]=(-q)T_{p,q+n},\\
 &[T_{m,n},T_{p,q}]=0.\label{W4(00)-deformed}
\end{split}
\end{equation}
 
To summarize, $\mathcal{W}(0,0;0,0)$ can be deformed  to the four parameter family ${\cal W}(a,b;\bar{a},\bar{b})$ or (exclusively) by three independent formal deformations parametrized by $\nu, \bar\nu, H_0$. Also, if one chooses to move in $(\bar{a},\bar{b})$ or $(a,b)$ planes, we are left with $\bar{\nu}$ or $\nu$ deformations, respectively, \emph{cf.} the  $\mathcal{W}(a,b;0,0)$ case discussed above.

\paragraph{Deformations of \texorpdfstring{$\mathcal{W}(0,-1;0,0)$}{u(1)Kac-Moody-bms3}. }
The next special case we study is $\mathcal{W}(0,-1;0,0)$:
\begin{equation} 
\begin{split}
    &[\mathcal{L}_{m},\mathcal{L}_{n}]=(m-n)\mathcal{L}_{m+n},\\
    &[\bar{\mathcal{L}}_{m},\bar{\mathcal{L}}_{n}]=(m-n)\bar{\mathcal{L}}_{m+n},\\
    &[\mathcal{L}_{m},\bar{\mathcal{L}}_{n}]=0,\\
    &[\mathcal{L}_{m},T_{p,q}]=(m-p)T_{p+m,q},\\
 &[\bar{\mathcal{L}}_{n},T_{p,q}]=(-q)T_{p,q+n},\\
 &[T_{m,n},T_{p,q}]=0,\label{u(1)-bms3}
\end{split}
\end{equation}
which is obtained from \eqref{W4-algebra} when we put $a=\bar{a}=\bar{b}=0$ and $b=-1$. As discussed $\mathcal{W}(0,-1;0,0)$ can be considered as combination of a $\mathfrak{u}(1)$ Kac-Moody algebra (on the right sector) and a \bmst\ (on the left sector). The global part of $\mathcal{W}(0,-1;0,0)$ is $\mathfrak{iso}(2,1)\oplus \mathfrak{sl}(2,\mathbb{R})$ spanned respectively by $\mathcal{L}_{r}, T_{r,0}$ and $\bar{\mathcal{L}}_{\bar{r}}$, $ r,{\bar{r}}=\pm1,0$.

Inspired by the discussions of previous subsection and recalling the results of \cite{Parsa:2018kys} for deformations of \bmst, we expect to be able to turn on a $T_{0,m+n}$ deformation in $[\bar{\mathcal{L}}_{m},\bar{\mathcal{L}}_{n}]$ and also be able to deform the ideal part; this is of course besides deforming by moving in the $(a,b;\bar{a},\bar{b})$ parameter space. The two allowed deformations are hence
\begin{equation} 
\begin{split}
 & [\mathcal{L}_{m},\mathcal{L}_{n}]=(m-n)\mathcal{L}_{m+n}, \\
 & [\bar{\mathcal{L}}_{m},\bar{\mathcal{L}}_{n}]=(m-n)\bar{\mathcal{L}}_{m+n}+\bar{\nu}\ (m-n)T_{0,m+n},\\
 &[{\mathcal{L}}_{m},\bar{\mathcal{L}}_{n}]=0,\\
 &[\mathcal{L}_{m},T_{p,q}]=(m-p)T_{p+m,q},\\
 &[\bar{\mathcal{L}}_{n},T_{p,q}]=(-q)T_{p,q+n},\\
 &[T_{m,n},T_{p,q}]=\varepsilon(m-p)T_{m+p,n+q}.
\end{split}\label{W4(0-100)-deformed}
\end{equation}
One can readily verify that the above deformations are formal. 

For the $\bar{\nu}=0$ the global part of the above algebra is $\mathfrak{sl}(2,\mathbb{R})\oplus \mathfrak{sl}(2,\mathbb{R}) \oplus \mathfrak{sl}(2,\mathbb{R})$, which is generated by ${\cal L}_r-\frac{1}{\varepsilon}T_{r,0}, \frac{1}{\varepsilon}T_{r,0}, \bar{\mathcal{L}}_{\bar r}$, $(r,\bar{r}=0,\pm 1)$. The first two $\mathfrak{sl}(2,\mathbb{R})$'s may be viewed as $\mathfrak{so}(2,2)$, the isometry of  AdS$_{3}$ space and the last $\mathfrak{sl}(2,\mathbb{R})$ factor as an ``internal symmetry'' for the AdS$_3$ space. One can then observe that the above algebra for $\bar{\nu}=0$ has a $\mathfrak{witt}\oplus \mathfrak{witt}\oplus \mathfrak{witt}$ subalgebra generated by ${\cal L}_n-\frac{1}{\varepsilon}T_{n,0}, \frac{1}{\varepsilon}T_{n,0}, \bar{\mathcal{L}}_{n}$. This latter has been studied as deformation of Maxwell algebra \cite{salgado2014so,gomis2009deformations, Concha:2018zeb,Caroca:2017onr}. Moreover, this algebra has a $\mathfrak{witt}\oplus \mathfrak{u}(1)$ Kac-Moody subalgebra generated by  ${\cal L}_n,  \bar{\mathcal{L}}_{n}, T_{0,n}$.

To summarize, one can deform $\mathcal{W}(0,-1;0,0)$ to a generic $\mathcal{W}(a,b;\bar{a},\bar{b})$ by moving in $(a,b;\bar{a},\bar{b})$ plane, or by turning on  $\nu$ or, exclusively, $\varepsilon(m-p)$ deformations.\footnote{Note that while $\nu$ and $\varepsilon$ deformations can be turned on simultaneously at infinitesimal level, they cannot both be elevated to a formal deformation at the same time.} If we move in $(\bar{a},\bar{b})$ plane we cannot turn on $\nu$ or $\varepsilon$ deformations and if we move in $(a,b)$ plane we cannot turn on $\varepsilon$ while $\nu$ deformation is possible. 
\section{Deformation of centrally extended \texorpdfstring{$\mathcal{W}(a,b;\bar{a},\bar{b})$}{w4} algebra, \texorpdfstring{$\widehat{\mathcal{W}}(a,b;\bar{a},\bar{b})$}{hat-w4} }\label{sec:6}

Global central extensions (which in short are usually called central extensions) of an algebra $\mathfrak{g}$ are classified by its second real cohomology ${\cal H}^2(\mathfrak{g};\mathbb{R})$. Central extensions may hence be viewed as a special class of deformations  given by Gel'fand-Fucks 2-cocycles \cite{gel1969cohomologies}. One may show, following analysis of \cite{Barnich:2011ct}, that $\mathcal{W}(a,b;\bar{a},\bar{b})$ for generic values of the parameters admits two independent central extensions which may be associated with deforming the algebra by two independent unit elements added to the algebra. The centrally extended ${\cal W}$-algebra which will be  denoted by $\widehat{\mathcal{W}}(a,b;\bar{a},\bar{b})$ is given by:
\begin{equation} 
\begin{split}
    &[\mathcal{L}_{m},\mathcal{L}_{n}]=(m-n)\mathcal{L}_{m+n}+\frac{C_{\mathcal{L}}}{12}m^{3}\delta_{m+n,0},\\
    &[\bar{\mathcal{L}}_{m},\bar{\mathcal{L}}_{n}]=(m-n)\bar{\mathcal{L}}_{m+n}+\frac{C_{\bar{\mathcal{L}}}}{12}m^{3}\delta_{m+n,0},\\
    &[\mathcal{L}_{m},\bar{\mathcal{L}}_{n}]=0,\\
    &[\mathcal{L}_{m},T_{p,q}]=-(p+bm+a)\ T_{p+m,q},\\
 &[\bar{\mathcal{L}}_{n},T_{p,q}]=-(q+\bar{b}n+\bar{a})\ T_{p,q+n},\\
 &[T_{m,n},T_{p,q}]=0,\label{central-w4}
\end{split}
\end{equation}
where $C_{\mathcal{L}}$ and $C_{\bar{\mathcal{L}}}$ are central charges. Algebras with different nonzero values of central charges $C_{\mathcal{L}}$ and $C_{\bar{\mathcal{L}}}$, are cohomologous, i.e they are isomorphic to each other. As we can see for the generic $a,b,\bar{a}$ and $\bar{b}$, in the centrally extended $\mathcal{W}(a,b;\bar{a},\bar{b})$ algebra the $\mathfrak{witt}$ subalgebras are turned to two Virasoro algebras and other commutators are untouched. 

As in the case of $W(a,b)$ algebras discussed in \cite{gao2011low, Parsa:2018kys}, there may be special points in the $(a,b,\bar{a},\bar{b})$ parameter space  which admit other central terms.
 As the first case let us consider \bmsf$={\cal W}(\frac{-1}{2},\frac{-1}{2},\frac{-1}{2},\frac{-1}{2})$.  This algebra  admits only two independent central terms in its two $\mathfrak{witt}$ algebras \cite{Barnich:2011ct}.

Recalling that in the $W(a,b)$ case there is a possibility of a central extension in ${\cal L}, T$ sector $a=0, b=1$ case \cite{gao2011low}, we examine if there is  a possibility of central extension in $[\mathcal{L}_{m},T_{p,q}]$ or $[\bar{\mathcal{L}}_{m},T_{p,q}]$ for specific values of $a,b,\bar{a}, \bar{b}$ parameters. Explicitly, consider
{
\begin{equation}
    [\mathcal{L}_{m},T_{p,q}]=-(a+bm+p)T_{m+p,q}+f(m,p)\delta_{q,0},
\end{equation}
and 
\begin{equation}
[\bar{\mathcal{L}}_{n},T_{p,q}]=-(\bar{a}+\bar{b}n+p)T_{p,n+q}+\bar{f}(n,q)\delta_{p,0},
\end{equation}
where $f(m,n)$ and $\bar{f}(m,n)$ are arbitrary functions. $[\mathcal{L}_{m},[\mathcal{L}_{n},T_{p,q}]]+cyclic\,\,permutations=0$ and $[\bar{\mathcal{L}}_{m},[\bar{\mathcal{L}}_{n},T_{p,q}]]+cyclic\,\,permutations=0$ Jacobi relations lead to 
\begin{equation}
   -(a+bn+p)f(m,p+n)+(a+bm+p)f(n,p+m)+(n-m)f(m+n,p)=0,\label{central-ab-LT}
\end{equation}
and 
\begin{equation}
   -(\bar{a}+\bar{b}n+p)\bar{f}(m,p+n)+(\bar{a}+\bar{b}m+p)\bar{f}(n,p+m)+(n-m)\bar{f}(m+n,p)=0.\label{central-bar-ab-LT}
\end{equation}
Let us now examine the equations for $m$ (or $n$)  or $p$ equal to zero. For $p=0$ \eqref{central-ab-LT}  yields
\begin{equation}
   -(a+bn)f(m,n)+(a+bm)f(n,m)+(n-m)f(m+n,0)=0.
\end{equation}
We can consider two cases, either $f$ is symmetric $f(m,n)=f(n,m)$, or it is antisymmetric $f(m,n)=-f(n,m)$. For the symmetric case we get $ bf(m,n)=f(m+n,0)$. So, either $b=0$ which leads to $f(m+n,0)=0$, or $b\neq 0$ for which $f(m,n)=\frac{1}{b}f(m+n,0)=F(m+n)$. For the former one learns that the only solution is $f(m,n)=m^{2}\delta_{m+n,0}$ as we expected for the $u(1)$ Kac-Moody algebra, \emph{cf.} \cite{Parsa:2018kys}. Plugging the solution $f(m,n)=F(m+n)$ into \eqref{central-ab-LT} restricts us to $b=1$ and arbitrary $a$.

For the antisymmetric $f(m,n)$, putting $p=0$ in \eqref{central-ab-LT} one finds 
\begin{equation}
   (2a+b(m+n))f(n,m)=(m-n)f(m+n,0),
\end{equation}
which for $m=0$ yields, either $b=-1, a=0$ or $f(n,0)=0$. One may then examine these two possibilities in \eqref{central-ab-LT} to find that $b=-1, a=0$ and that $f(m,n)=m^3\delta_{m+n,0}$ is the only non-trivial solution. This latter is of course expected recalling the \bmst\ analysis of \cite{Parsa:2018kys}. Similar analysis goes through for \eqref{central-bar-ab-LT}.} To summarize so far, the $f,\bar{f}$ type central terms are allowed only for 
    $a=b=0, f(m,n)=m^2\delta_{m+n,0}$; 
$b=1, a=arbitrary, f(m,n)=F(m+n)$; 
$b=-1, a=0, f(m,n)=m^3\delta_{m+n,0}$.

We should now verify if the central terms in special points  obtained above satisfy the Jacobi 
$[\mathcal{L}_{r},[\bar{\mathcal{L}}_{s},T_{m,n}]]+cyclic\,\,permutations=0$.
For generic $a,b,\bar{a}$ and $\bar{b}$ one obtains
\begin{equation}
    (a+br+p)\bar{f}(s,q)-(\bar{a}+\bar{b}s+q)f(r,p)=0. \label{C-f-barf-LLT}
\end{equation}
For the special point $a=b=\bar{a}=\bar{b}=0$ corresponding to  ${\cal W}(0,0;0,0)$ algebra,
we obtained $\bar{f}(m,n)=f(m,n)=m^{2}\delta_{m+n,0}$  which does not fulfill \eqref{C-f-barf-LLT}. The next point is ${\cal W}(0,-1;0,0)$, which can be viewed as combination of \bmst\ and $u(1)$ Kac-Moody, we obtained $f(m,n)=m^{3}\delta_{m+n,0}$ and $\bar{f}(m,n)=m^{2}\delta_{m+n,0}$. This too, does not satisfy \eqref{C-f-barf-LLT}. One therefore concludes that $\mathcal{W}(0,0;0,0)$ and ${\cal W}(0,-1;0,0)$ do not admit a central term in its $[\mathcal{L}_{m},T_{p,q}]$ or $[\bar{\mathcal{L}}_{m},T_{p,q}]$ commutators. But one can consider $\mathcal{W}(0,1;0,0)$ and $\mathcal{W}(0,1;0,-1)$ which admit central terms as $\bar{f}(m,n)=m^{2}\delta_{m+n,0}$ and $\bar{f}(m,n)=m^{3}\delta_{m+n,0}$ respectively. For the special case $\mathcal{W}(0,1;0,1)$ one learns that it admits two independent central terms as $f(m,n)=(\alpha +\beta m)\delta_{m+n,0}$ and $\bar{f}(m,n)=(\bar{\alpha}+\bar{\beta }m)\delta_{m+n,0}$.

The specific case ${\cal W}(0,0;a,b)$ algebra which can be seen as combination of  $W(a,b)$ and $u(1)$ Kac-Moody algebras may also admit central terms in its ideal part which can be parametrized as
\begin{equation}
    [T_{m,n},T_{p,q}]= c_1m\delta_{m+p,0}\delta_{n,q}+c_2n\delta_{m,p}\delta_{n+q,0}.
\end{equation}
This structure guarantees antisymmetry w.r.t. $m\leftrightarrow p$ and $n\leftrightarrow q$. 
The Jacobi $[\mathcal{L}_{r},[T_{m,n},T_{p,q}]]+cyclic\,\,permuutation=0$ leads to
\begin{equation}
   -p\big((\bar{a}+\bar{b}r+q)\delta_{n,q+r}+(\bar{a}+\bar{b}r+n)\delta_{q,n+r}\big)=0,
\end{equation}
which cannot be satisfied for any values of $\bar{a}$ and $\bar{b}$. Therefore, one finds that $\mathcal{W}(0,0;\bar{a},\bar{b})$ admits only central terms in its two Witt algebras (unless the case $\bar{a}=0$ and $\bar{b}=1$), just as \bmsf\ and $\mathcal{W}(a,b;\bar{a},\bar{b})$ for generic $a,b,\bar{a}$ and $\bar{b}$. 

\subsection{Most general deformations of centrally extended \texorpdfstring{$\mathcal{W}(a,b;\bar{a},\bar{b})$}{w4} algebra}

Let us now consider the most general deformations of the $
\widehat{\mathcal{W}}(a,b;\bar{a},\bar{b})$ algebra. As we checked in the previous part $
\widehat{\mathcal{W}}(a,b;\bar{a},\bar{b})$ admits two nontrivial central charges in its two Witt algebras. We start with this algebra and schematically consider its most general deformations as
\begin{equation} 
\begin{split}
 & [\mathcal{L},\mathcal{L}]=\mathcal{L}+C_{\mathcal{L}}+hT+X, \\
 & [\bar{\mathcal{L}},\bar{\mathcal{L}}]=\bar{\mathcal{L}}+\bar{C}_{\bar{\mathcal{L}}}+\bar{h}T+\bar{X},\\
 &[{\mathcal{L}},\bar{\mathcal{L}}]=HT+U,\\
 &[\mathcal{L},T]=T+KT+fL+g\bar{L}+Y,\\
 &[\bar{\mathcal{L}},T]=T+\bar{K}T+\bar{f}L+\bar{g}\bar{L}+\bar{Y},\\
 &[T,T]=GT+AL+B\bar{L}+Z.
\end{split}\label{most general-hatW4-algebra}
\end{equation}
in which we dropped the indices of generators and arguments of functions to simplify the notation. The functions $X$, $U$, $Y$ and $Z$ and their analogues in barred sector, are deformations by terms with coefficients in unit generators (central terms). As we have seen in the case of $W(a,b)$ algebra which has been studied in \cite{Parsa:2018kys}, the Jacobi analysis leads to two different family of relations. The first family is exactly the same as relations analyzed in the previous section for the functions related to non-central parts while the second set of relations include linear combinations of central and non-central functions. 
In this way, for the generic values of $a,b,\bar{a}$ and $\bar{b}$  we obtained that the only nontrivial solutions are $K(m,n)=\alpha+\beta m$ and $\bar{K}(m,n)=\bar{\alpha}+\bar{\beta}m$ which lead to a new $\widehat{\mathcal{W}}(a,b;\bar{a},\bar{b})$ with shifts in the four parameters and none of the other functions, central or non-central, cannot be turned on. So, one concludes that the family of $\widehat{\mathcal{W}}(a,b;\bar{a},\bar{b})$ algebras are rigid (or stable), in the sense that it can just be deformed to another $\widehat{\mathcal{W}}(a,b;\bar{a},\bar{b})$ in the same family. As we discussed, however, there are  special points in the space $(a,b,\bar{a},\bar{b})$ which can admit some other deformations. Now, we are going to review the results of the most general 
deformations of $\widehat{\mathcal{W}}(a,b;\bar{a},\bar{b})$ in special points.  
 
 \paragraph{The most general deformations of \hbmsf.}
As the first case we consider $\widehat{\mathcal{W}}(\frac{-1}{2},\frac{-1}{2};\frac{-1}{2},\frac{-1}{2})$ which is the central extension of \bmsf\ denoted by \hbmsf . As mentioned, from the first family of relations, the only nontrivial functions are $K(m,n)=\alpha+\beta m$ and $\bar{K}(m,n)=\bar{\alpha}+\bar{\beta}m$ and other non-central functions are zero, as we have shown  in section \ref{sec:4}. The Jacobi $[\mathcal{L},[\bar{\mathcal{L}},T]]+cyclic\,\,permutation=0$ leads to $Y=\bar{Y}=0$. The Jacobi $[\bar{\mathcal{L}},[\mathcal{L},\mathcal{L}]]+cyclic\,\,permutation=0$ leads to $U=0$. The Jacobi $[\mathcal{L},[T,T]]+cyclic\,\,permutation=0$ leads to $Z=0$. The Jacobi $[\mathcal{L},[\mathcal{L},\mathcal{L}]]+cyclic\,\,permutation=0$ leads to $X(m)=m^{3}\delta_{m+n,0}$ and a similar result for $\bar{X}(m)$ which just lead to a shift of $C_{\mathcal{L}}$ and $C_{\bar{\mathcal{L}}}$. However, as mentioned, the algebras with different values of $C_{\mathcal{L}}$ and $C_{\bar{\mathcal{L}}}$ are cohomologous and isomorphic to each other. In this way, we have found that the most general deformations of \hbmsf$=\widehat{\mathcal{W}}(-1/2,-1/2;-1/2,-1/2)$  is $\widehat{\mathcal{W}}(a,b;\bar{a},\bar{b})$ with the shifted parameters. 

\subsection{Most general deformations of specific points in \texorpdfstring{$(a,b;\bar{a},\bar{b})$}{abab} space}
The next special point is $\widehat{\mathcal{W}}(0,0;0,0)$. We showed in the  previous section that non-central functions with nontrivial solutions are  $h(m,n)=constant$, $K(m,n)=\alpha+\beta m$ and the same result for $\bar{h}(m,n)$ and $\bar{K}(m,n)$ and $H(m,n)=H_{0}(\alpha+\beta m)(\bar{\alpha}+\bar{\beta }n)$. 
The Jacobi $[\mathcal{L},[\bar{\mathcal{L}},T]]+cyclic\,\,permutation=0$ leads to $Y=\bar{Y}=0$. The Jacobi $[\bar{\mathcal{L}},[\mathcal{L},\mathcal{L}]]+cyclic\,\,permutation=0$ leads also to $U=0$. The Jacobi $[\mathcal{L},[T,T]]+cyclic\,\,permutation=0$ leads to $Z=0$. The Jacobi $[\mathcal{L},[\mathcal{L},\mathcal{L}]]+cyclic\,\,permutation=0$ leads to $X(m)=m^{3}\delta_{m+n,0}$ and a similar result for $\bar{X}(m)$ which just shifts the value of $C_{\mathcal{L}}$ and $C_{\bar{\mathcal{L}}}$.
We hence recover exactly the same results as the infinitesimal single deformation case. 

The next special point is $\widehat{\mathcal{W}}_{4}(0,-1;0,0)$. Non-central functions with nontrivial solutions are  $\bar{h}(m,n)=constant$, $K(m,n)=\alpha+\beta m$, $\bar{K}(m,n)=\bar{\alpha}+\bar{\beta} n$ and $G(m,n;p,q)=(m-p)$ and other functions are zero. The Jacobi $[\mathcal{L},[\bar{\mathcal{L}},T]]+cyclic\,\,permutation=0$ leads to $Y=\bar{Y}=0$. The Jacobi $[\bar{\mathcal{L}},[\mathcal{L},\mathcal{L}]]+cyclic\,\,permutation=0$ leads to $U=0$. The Jacobi $[\mathcal{L},[\mathcal{L},\mathcal{L}]]+cyclic\,\,permutation=0$ leads to $X(m)=m^{3}\delta_{m+n,0}$ and a similar result for $\bar{X}(m)$. Finally, when we deform the ideal part as $[T_{m,n},T_{p,q}]=(m-p)T_{m+p,n+q}+Z(m,n;p,q)$, the Jacobi $[\mathcal{L}_{m},[T_{p,q},T_{r,s}]]+cyclic\,\,permutation=0$ leads to $Z(m,n;p,q)=0$. 

One may consider the specific subalgebra of $\widehat{\mathcal{W}}_{4}(0,-1;0,0)$ generated by $\mathcal{L}^1_n\equiv T_{n,0}, \mathcal{L}^2_n=\mathcal{L}_n- T_{n,0}, \mathcal{L}^3_n=\bar{\mathcal{L}}_n$ for which, $$[\mathcal{L}^a_n,\mathcal{L}^b_n]=\delta^{ab}\left( (n-m) \mathcal{L}^a_{n+m}+\frac{1}{12} C^a n^3\delta_{m+n,0}\right),\quad a,b=1,2,3.$$ For this subalgebra,  $Z(m,n;p,q)=m^{3}\delta_{m+p,0}\delta_{n,0}\delta_{q,0}$ is an allowed central extension (as well as a formal deformation), denoted by $C^a$ in the above algebra. This subalgebra hence admits three central charges and is therefore, direct sum of three Virasoro algebras.

The next specific point is $\widehat{\mathcal{W}}_{4}(0,1;0,0)$.  Non-central functions with nontrivial solutions are  $\bar{h}(m,n)=constant$, $K(m,n)=\alpha+\beta m$ and $\bar{K}(m,n)=\bar{\alpha}+\bar{\beta}n$. The Jacobi $[\bar{\mathcal{L}},[\mathcal{L},\mathcal{L}]]+cyclic\,\,permutation=0$ leads to $U=0$. The Jacobi $[\mathcal{L},[\mathcal{L},\mathcal{L}]]+cyclic\,\,permutation=0$ leads to $X(m)=m^{3}\delta_{m+n,0}$ and a similar result  for $\bar{X}(m)$. The Jacobi $[\mathcal{L},[T,T]]+cyclic\,\,permutation=0$ leads to $Z=0$. The Jacobi $[\mathcal{L},[\bar{\mathcal{L}},T]]+cyclic\,\,permutation=0$ yields $Y=0$. Unlike the previous cases, however, we obtained $\bar{Y}(m)=m^{2}\delta_{m+n,0}$. Although the latter is a formal deformation, when we turn on $\bar{K}$, $\bar{h}$ and $\bar{Y}$ simultaneously they cannot satisfy the Jacobi in higher order in deformation parameter, so they should be considered as independent formal deformations.

The next specific case is $\widehat{\mathcal{W}}_{4}(0,1;0,-1)$.  Non-central functions with nontrivial solutions are  $K(m,n)=\alpha+\beta m$ and $\bar{K}(m,n)=\bar{\alpha}+\bar{\beta}n$. The Jacobi $[\bar{\mathcal{L}},[\mathcal{L},\mathcal{L}]]+cyclic\,\,permutation=0$ leads to $U=0$,  $[\mathcal{L},[\mathcal{L},\mathcal{L}]]+cyclic\,\,permutation=0$  to $X(m)=m^{3}\delta_{m+n,0}$, and a similar result for $\bar{X}(m)$. The Jacobi $[\mathcal{L},[\bar{\mathcal{L}},T]]+cyclic\,\,permutation=0$ leads to $Y=0$ and $\bar{Y}(m)=m^{3}\delta_{m+n,0}$. As in the previous cases one can show that the functions $\bar{K}$ and $\bar{Y}$ cannot be turned on simultaneously, implying that we do not have formal deformations induced with both $\bar{K}$ and $\bar{Y}$. 

The next specific point is $\widehat{\mathcal{W}}_{4}(0,1;0,1)$. Non-central functions with nontrivial solutions are  $K(m,n)=\alpha+\beta m$ and $\bar{K}(m,n)=\bar{\alpha}+\bar{\beta}n$.The Jacobi $[\bar{\mathcal{L}},[\mathcal{L},\mathcal{L}]]+cyclic\,\,permutation=0$ leads to $U=0$, and $[\mathcal{L},[\mathcal{L},\mathcal{L}]]+cyclic\,\,permutation=0$ to $X(m)=m^{3}\delta_{m+n,0}$, and similarly for $\bar{X}(m)$. The Jacobi $[\mathcal{L},[\bar{\mathcal{L}},T]]+cyclic\,\,permutation=0$ leads to $Y(m)=(\tilde{\alpha} +\tilde{\beta} m)\delta_{m+n,0}$ and  $\bar{Y}(m)=(\tilde{\bar{\alpha}} +\tilde{\bar{\beta}} m)\delta_{m+n,0}$. One can show that the functions $K, Y$ and $\bar{K}, \bar{Y}$ cannot be turned on simultaneously.

  
 \section{ Cohomological consideration of \texorpdfstring{$\mathcal{W}(a,b;\bar{a},\bar{b})$}{w4}  algebra }\label{sec:7}
 
The direct and explicit verification of Jacobi identities for deformations may be presented in the language of algebraic cohomology. Following our discussions for  \bmst\ in \cite{Parsa:2018kys}, here we  study second adjoint cohomology of \bmsf\ and its central extension \hbmsf\ as well as the $\mathcal{W}(a,b;\bar{a},\bar{b})$ and its central extension $\widehat{\mathcal{W}}(a,b;\bar{a},\bar{b})$ for generic $a,b,\bar{a}$ and $\bar{b}$. 
The main tools to this cohomological analysis is the Hochschild-Serre spectral sequence  which has been reviewed in appendix \ref{appendix-B}. {As similar analysis for infinite dimensional Schr\"{o}dinger-Virasoro type algebras may be found in \cite{Unterberger:2011yya}.}

\subsection{Cohomological consideration of \texorpdfstring{$\mathfrak{bms}_{4}$}{BMS4}  algebra}

As we know $\mathcal{H}^{2}(\mathfrak{bms}_{4};\mathfrak{bms}_{4})$ classifies all infinitesimal deformations of $\mathfrak{bms}_{4}$ algebra which may be computed using 
the spectral sequence \eqref{E2-dec} and the long exact sequences \eqref{long-exact} {and \eqref{long-exact2}} . By the former we obtain  information about $\mathcal{H}^{2}(\mathfrak{bms}_{4};(\mathfrak{witt}\oplus\mathfrak{witt}))$ and $\mathcal{H}^{2}(\mathfrak{bms}_{4};\mathfrak{T})$ independently where $\mathfrak{T}$ and $(\mathfrak{witt}\oplus\mathfrak{witt})$ respectively denote the ideal part and the Witt subalgebra of the $\mathfrak{bms}_{4}$. Note that since $(\mathfrak{witt}\oplus\mathfrak{witt})$ algebra is not a \bmsf\ module {by the adjoint action,} $\mathcal{H}^{2}(\mathfrak{bms}_{4};(\mathfrak{witt}\oplus\mathfrak{witt}))$ is defined by the action used in the short exact sequence \eqref{short-exact-bms} below, as discussed in appendix \ref{appendix-B}. Note also that given the semi-direct sum structure of the \bmsf\ algebra \eqref{bms=witt+ideal}, one should not expect 
$\mathcal{H}^{2}(\mathfrak{bms}_{4};\mathfrak{bms}_{4})$ to be equal to $\mathcal{H}^{2}(\mathfrak{bms}_{4};(\mathfrak{witt}\oplus\mathfrak{witt})) \oplus \mathcal{H}^{2}(\mathfrak{bms}_{4};\mathfrak{T})$. Nonetheless, as we will see, direct analysis of $\mathcal{H}^{2}(\mathfrak{bms}_{4};(\mathfrak{witt}\oplus\mathfrak{witt}))$ and $ \mathcal{H}^{2}(\mathfrak{bms}_{4};\mathfrak{T})$  {may carry information about the} structure of $\mathcal{H}^{2}(\mathfrak{bms}_{4};\mathfrak{bms}_{4})$.    

To this end, we follow the Hochschild-Serre spectral sequence method (\emph{cf}. appendix \ref{appendix-B}) and consider the following short exact sequence of $\mathfrak{bms}_{4}$ algebra 
\begin{equation}
    0\longrightarrow \mathfrak{T}_{ab}\longrightarrow \mathfrak{bms}_4 \longrightarrow \mathfrak{bms}_4/\mathfrak{T}_{ab}\cong (\mathfrak{witt}\oplus\mathfrak{witt})\longrightarrow 0,\label{short-exact-bms}
\end{equation}
where $\mathfrak{T}_{ab}$ is the abelian ideal of the $\mathfrak{bms}_4$ algebra which is spanned by $T_{m,n}$ generators.
As  in the case of $\mathfrak{bms}_{3}$ \cite{Parsa:2018kys}, we compute $\mathcal{H}^{2}(\mathfrak{bms}_{4};\mathfrak{witt}\oplus\mathfrak{witt})$ and $\mathcal{H}^{2}(\mathfrak{bms}_{4};\mathfrak{T}_{ab})$ separately. 

\paragraph{Computation of  $\mathcal{H}^{2}(\mathfrak{bms}_{4};\mathfrak{T})$.}
As is reviewed in appendix \ref{appendix-B} and from \eqref{E2-dec} and \eqref{dec2} one can compute the $\mathcal{H}^{2}(\mathfrak{bms}_{4};\mathfrak{T})$ as 
\begin{equation}
\begin{split}
    \mathcal{H}^{2}(\mathfrak{bms}_{4};\mathfrak{T})&=\oplus_{p+q=2}E_{2;\mathfrak{T}}^{p,q}=E_{2;\mathfrak{T}}^{2,0}\oplus E_{2;\mathfrak{T}}^{1,1}\oplus E_{2;\mathfrak{T}}^{0,2}\\
    &=\mathcal{H}^{2}((\mathfrak{witt}\oplus\mathfrak{witt});\mathcal{H}^{0}(\mathfrak{T};\mathfrak{T}))\oplus \mathcal{H}^{1}((\mathfrak{witt}\oplus\mathfrak{witt});\mathcal{H}^{1}(\mathfrak{T};\mathfrak{T}))\cr &\oplus \mathcal{H}^{0}((\mathfrak{witt}\oplus\mathfrak{witt});\mathcal{H}^{2}(\mathfrak{T};\mathfrak{T})),\label{bmsT}
\end{split}
\end{equation}
where the subscript $\mathfrak{T}$ in $E_{2;\mathfrak{T}}^{p,q}$ means we are computing $\mathcal{H}^{2}(\mathfrak{bms}_{4};\mathfrak{T})$.
We compute the  three terms  above separately. 
$\mathcal{H}^{2}((\mathfrak{witt}\oplus\mathfrak{witt});\mathcal{H}^{0}(\mathfrak{T};\mathfrak{T}))$ contains $\mathcal{H}^{0}(\mathfrak{T};\mathfrak{T})$ which by the definition (3.14) of \cite{Parsa:2018kys} (or \cite{fuks2012cohomology}) and the fact that the action of $\mathfrak{T}$ on $\mathfrak{T}$ is trivial, one concludes that $\mathcal{H}^{0}(\mathfrak{T};\mathfrak{T})=\mathfrak{T}$ then $\mathcal{H}^{2}((\mathfrak{witt}\oplus\mathfrak{witt});\mathcal{H}^{0}(\mathfrak{T};\mathfrak{T}))=\mathcal{H}^{2}((\mathfrak{witt}\oplus\mathfrak{witt});\mathfrak{T})$. On the other hand, by the direct computations in subsection \ref{Witt-deformations-sec} we have shown that two Witt subalgebras in \bmsf\ do not admit any nontrivial deformation with terms by coefficients in $T$. So one concludes that $\mathcal{H}^{2}((\mathfrak{witt}\oplus\mathfrak{witt});\mathfrak{T})=0$. 
Therefore the two first commutators in \eqref{bms4} remains intact by deformation procedure, in accord with results of subsection \ref{Witt-deformations-sec}.

Next we  analyze $\mathcal{H}^{1}((\mathfrak{witt}\oplus\mathfrak{witt});\mathcal{H}^{1}(\mathfrak{T};\mathfrak{T}))$. It  is constructed by 1-cocycle $\varphi_{1}$ which is defined as a function $ \varphi_{1}: (\mathfrak{witt}\oplus\mathfrak{witt})\longrightarrow (\mathfrak{T};\mathfrak{T})\stackrel{\tilde\varphi}{\longrightarrow} \mathfrak{T}$, where here $\tilde\varphi$ is a 1-cocycle used in the definition of $\mathcal{H}^{1}(\mathfrak{T};\mathfrak{T})$.
The expression of $\varphi_{1}(\mathcal{L}_{m},\bar{\mathcal{L}}_{m})(T_{p,q})$ can hence be expanded in terms of $T$'s as $\varphi_{1}(\mathcal{L}_{m},\bar{\mathcal{L}}_{m})(T_{p,q})=\tilde{K}(m,p)T_{m+p,q}+ \tilde{\bar{K}}(m,q)T_{p,m+q},\label{Cjpp}$
where $\tilde{K}(m,p)$ and $\tilde{\bar{K}}(m,q)$ are arbitrary functions. 
The deformation of $[\mathcal{L},T]$ part corresponding to $\varphi_{1}$ is $[\mathcal{L}_m,T_{p,q}]=(\frac{m+1}{2}-p)T_{m+p,q}+\tilde{K}(m,p){T}_{m+p,q}$.
The Jacobi identity for the above bracket imposes restraints on $\tilde{K}(m,n)$ exactly like the ones on $K(m,n)$ in \eqref{eq-K}, so one finds the same result as $\tilde{K}(m,n)=\alpha +\beta m$ and the same result is obtained for $\tilde{\bar{K}}(m,n)$.

The last term we study is  $\mathcal{H}^{0}((\mathfrak{witt}\oplus\mathfrak{witt});\mathcal{H}^{2}(\mathfrak{T};\mathfrak{T}))$. We use the definition of $\mathcal{H}^{0}$ 
\begin{equation}
  \mathcal{H}^{0}((\mathfrak{witt}\oplus\mathfrak{witt});\mathcal{H}^{2}(\mathfrak{T};\mathfrak{T}))=\{ \psi  \in \mathcal{H}^{2}(\mathfrak{T};\mathfrak{T})| (\mathcal{L}\oplus\bar{\mathcal{L}})\circ \psi=0 ,\,\,\, \forall\,\, \mathcal{L}\,\, \text{and}\,\, \bar{\mathcal{L}}\in (\mathfrak{witt}\oplus\mathfrak{witt}) \},\label{H0T}
\end{equation}
 where $\psi$ is a $T$-valued  2-cocycle. 
 The action ``$\circ$'' of $\mathcal{L}$ on a 2-cocycle $\psi$ is defined as \cite{MR0054581}
\begin{equation}
  (\mathcal{L}_{r}\circ \psi)(T_{m,n},T_{p,q})=[\mathcal{L}_{r},\psi(T_{m,n},T_{p,q})]-\psi([\mathcal{L}_{r},T_{m,n}],T_{p,q})-\psi(T_{m,n},[\mathcal{L}_{r},T_{p,q}]),\label{ideal-jacobi}
\end{equation}
Expanding $\psi$ in terms of $T$s as $\psi(T_{m,n},T_{p,q})=G(m,n;p,q)T_{m+p,n+q}$, we get the same relation as (\ref{TTT}) which has the solution $G(m,n;p,q)=0$. The same relation can be obtained by $\bar{\mathcal{L}}$. 
The above discussion leads to 
\begin{equation}
    \mathcal{H}^{2}(\mathfrak{bms}_{4};\mathfrak{T})=\mathcal{H}^{1}((\mathfrak{witt}\oplus\mathfrak{witt});\mathcal{H}^{1}(\mathfrak{T};\mathfrak{T})),\label{bmsT2}
\end{equation}
which means that turning on deformations with coefficients in $\mathfrak{T}$, we can only  deform the $[\mathcal{L},T]$ part by $\tilde{K}(m,n)$. This is exactly in agreement of our results of direct and explicit calculations in section \ref{sec:3}.

 \paragraph{Computation of  $\mathcal{H}^{2}(\mathfrak{bms}_{4};(\mathfrak{witt}\oplus\mathfrak{witt}))$.}
One can expand the latter as
\begin{equation}
\begin{split}
     \mathcal{H}^{2}(\mathfrak{bms}_{4};(\mathfrak{witt}\oplus\mathfrak{witt}))&=\oplus_{p+q=2}E_{2;(\mathfrak{w}\oplus\mathfrak{w})}^{p,q}=E_{2;(\mathfrak{w}\oplus\mathfrak{w})}^{2,0}\oplus E_{2;(\mathfrak{w}\oplus\mathfrak{w})}^{1,1}\oplus E_{2;(\mathfrak{w}\oplus\mathfrak{w})}^{0,2}\\
    &=\mathcal{H}^{2}((\mathfrak{witt}\oplus\mathfrak{witt});\mathcal{H}^{0}(\mathfrak{T};(\mathfrak{witt}\oplus\mathfrak{witt})))\cr& \oplus \mathcal{H}^{1}((\mathfrak{witt}\oplus\mathfrak{witt});\mathcal{H}^{1}(\mathfrak{T};(\mathfrak{witt}\oplus\mathfrak{witt})))\cr &\oplus \mathcal{H}^{0}((\mathfrak{witt}\oplus\mathfrak{witt});\mathcal{H}^{2}(\mathfrak{T};(\mathfrak{witt}\oplus\mathfrak{witt}))),\label{bms-LbarL}
\end{split}
\end{equation}
where the subscript $(\mathfrak{w}\oplus\mathfrak{w})$ denotes we are computing $  \mathcal{H}^{2}(\mathfrak{bms}_{4};(\mathfrak{witt}\oplus\mathfrak{witt}))$.

To compute  $\mathcal{H}^{2}((\mathfrak{w}\oplus\mathfrak{w});\mathcal{H}^{0}(\mathfrak{T};(\mathfrak{witt}\oplus\mathfrak{witt})))$, we recall that the action  of $\mathfrak{T}$ on $(\mathfrak{witt}\oplus\mathfrak{witt})$ (which is induced via the short exact sequence  (\ref{short-exact-bms})),  is trivial and  hence $\mathcal{H}^{0}(\mathfrak{T};(\mathfrak{witt}\oplus\mathfrak{witt}))\cong (\mathfrak{witt}\oplus\mathfrak{witt})$. We then conclude 
\be\label{H2-witt-witt}
\mathcal{H}^{2}((\mathfrak{witt}\oplus\mathfrak{witt});\mathcal{H}^{0}(\mathfrak{T};(\mathfrak{witt}\oplus\mathfrak{witt})))\cong  \mathcal{H}^{2}((\mathfrak{witt}\oplus\mathfrak{witt});(\mathfrak{witt}\oplus\mathfrak{witt}))\cong 0,
\ee 
where in the last step we used the fact that $(\mathfrak{witt}\oplus\mathfrak{witt})$ algebra is rigid \cite{Parsa:2018kys}.

Next, we consider the second term in \eqref{bms-LbarL}, $\mathcal{H}^{1}((\mathfrak{witt}\oplus\mathfrak{witt});\mathcal{H}^{1}(\mathfrak{T};(\mathfrak{witt}\oplus\mathfrak{witt})))$ which is constructed by 1-cocycle $\varphi_{2}$ as $\varphi_{2}: (\mathfrak{witt}\oplus\mathfrak{witt})\longrightarrow \mathcal{H}^{1}(\mathfrak{T};(\mathfrak{witt}\oplus\mathfrak{witt}))$.
A similar analysis as the previous case tells us that $\varphi_2$ deforms the commutator $[\mathcal{L},T]$ part as $[\mathcal{L}_{m},T_{p,q}]=(\frac{m+1}{2}-p)T_{m+p,q}+f(m,p)\mathcal{L}_{p+m-1}\delta_{q,0}+g(m)\bar{\mathcal{L}}_{q-1}\delta_{m+p,0}$ (and a similar relation can be obtained for $[\bar{\mathcal{L}},T]$). Recalling the arguments of previous section, 
one concludes 
\be
\mathcal{H}^{1}((\mathfrak{witt}\oplus\mathfrak{witt});\mathcal{H}^{1}(\mathfrak{T};(\mathfrak{witt}\oplus\mathfrak{witt})))=0.
\ee
This means that the $[\mathcal{L},T]$ and $[\bar{\mathcal{L}},T]$ commutators cannot be deformed by  terms with coefficients {in} $\mathcal{L}$ and $\bar{\mathcal{L}}$.

We finally compute the last term in \eqref{bms-LbarL}, $\mathcal{H}^{0}((\mathfrak{witt}\oplus\mathfrak{witt});\mathcal{H}^{2}(\mathfrak{T};(\mathfrak{witt}\oplus\mathfrak{witt})))$.  One can repeat the procedure exactly the same as the previous case to get 
\begin{equation}
  (\mathcal{L}_{m}\circ \psi)(T_{p,q},T_{r,s})=[\mathcal{L}_{m},\psi(T_{p,q},T_{r,s})]-\psi([\mathcal{L}_{m},T_{p,q}],T_{r,s})-\psi(T_{r,s},[\mathcal{L}_{m},T_{p,q}]),\label{hjacobi-1}
\end{equation}
and
\begin{equation}
  (\bar{\mathcal{L}}_{m}\circ \psi)(T_{p,q},T_{m,n})=[\bar{\mathcal{L}}_{m},\psi(T_{p,q},T_{m,n})]-\psi([\bar{\mathcal{L}}_{m},T_{p,q}],T_{m,n})-\psi(T_{m,n},[\bar{\mathcal{L}}_{m},T_{p,q}]),\label{hjacobi-2}
\end{equation}
with $\psi(T_{p,q},T_{m,n})=A(m,n;p,q)\mathcal{L}_{m+p-1}+B(m,n;p,q)\bar{\mathcal{L}}_{n+q-1}$ where $A(m,n;p,q)$ and $B(m,n;p,q)$ are arbitrary antisymmetric functions. Inserting the latter into \eqref{hjacobi-1} and \eqref{hjacobi-2}  we get the same relation as \eqref{B-eq0} and its analogue for $A$, leading to $A(m,n;p,q)=B(m,n;p,q)=0$. Therefore,  $\mathcal{H}^{0}((\mathfrak{witt}\oplus\mathfrak{witt});\mathcal{H}^{2}(\mathfrak{T};(\mathfrak{witt}\oplus\mathfrak{witt})))=0$ for  \bmsf\ case. This is in contrast to the \bmst\ case \cite{Parsa:2018kys}.

As summary of the above discussions one concludes that 
\be\mathcal{H}^{2}(\mathfrak{bms}_{4},\mathfrak{bms}_{4})=\mathcal{H}^{1}((\mathfrak{witt}\oplus\mathfrak{witt});\mathcal{H}^{1}(\mathfrak{T};\mathfrak{T})).\label{bms-witt}
\end{equation}
That is,  deformations of $\mathfrak{bms}_{4}$ are just those that deform the $[\mathcal{L},T]$ and $[\bar{\mathcal{L}},T]$ commutators by terms with coefficients in $T$. In other words, \bmsf\ algebra can only be deformed into \wfab, in accord with our direct Jacobi identity analysis of previous section.

\subsection{Cohomological consideration for \texorpdfstring{$\widehat{\mathfrak{bms}}_{4}$}{BMS4}  algebra}

As discussed in \cite{Oblak:2016eij,Parsa:2018kys} for a given algebra $\mathfrak{g}$, (1) central extensions are classified by Gel'fand-Fucks 2-cocycles, or by $\mathcal{H}^{2}(\mathfrak{g}; \mathbb{R})$ and, (2)
to deal with the central extensions in the cohomological analysis we need to extend the algebra by adding unit elements, one unit element for each possible central term, explicitly, we need to consider $\hat{\mathfrak{g}}=\mathfrak{g}\oplus \mathfrak{u}(1)\oplus\cdots \oplus \mathfrak{u}(1)$, where the number of $\mathfrak{u}(1)$ factors is equal to the number of independent commutators in the algebra, i.e. for \bmst\ it is three and for \bmsf\ it is six. Of course closure condition may not allow to turn on all these central terms.\footnote{{For instance, the Jacobi identity allows \bmst\ to have two central terms in the $[\mathcal{J},\mathcal{J}]$ and $[\mathcal{J},\mathcal{P}]$ commutators, for the $\mathfrak{KM}_{\mathfrak{u}(1)}$ we can also have central term in $[\mathcal{P},\mathcal{P}]$ commutators, altogether three central extensions \cite{gao2011low, Parsa:2018kys}. The \bmsf, however, just admits two central terms in its two Witt subalgebras.}} In our analysis, we adopt the viewpoint that each central extension is a deformation in the $\mathfrak{u}(1)$ extended algebra and that turning on a central charge is like a deformation in $\hat{\mathfrak{g}}$. Therefore, we need not first study $\mathcal{H}^{2}(\mathfrak{g}; \mathbb{R})$ and then analyze the deformation, we may directly focus on $\mathcal{H}^{2}(\hat{\mathfrak{g}}; \hat{\mathfrak{g}})$. 

{As in the $\mathfrak{bms}_{4}$ case, $(\mathfrak{vir}\oplus\mathfrak{vir})$ is not a $\widehat{\mathfrak{bms}}_{4}$ module by the adjoint action and $\mathcal{H}^{2}(\widehat{\mathfrak{bms}}_{4};(\mathfrak{vir}\oplus\mathfrak{vir}))$ is defined by the action induced from the short exact sequence \eqref{short-exact-vir-vir}, as was discussed, we should not expect
$\mathcal{H}^{2}(\widehat{\mathfrak{bms}}_{4};\widehat{\mathfrak{bms}}_{4})$ to be equal to $\mathcal{H}^{2}(\widehat{\mathfrak{bms}}_{4};(\mathfrak{vir}\oplus\mathfrak{vir}))\oplus \mathcal{H}^{2}(\widehat{\mathfrak{bms}}_{4};\mathfrak{T}_{ab})$.} 
As in the \bmsf\ case, computation of each of these two factors would be helpful in computing the former.
To this end we employ the Hochschild-Serre spectral sequence theorem. The $\widehat{\mathfrak{bms}}_{4}$ has semi-direct sum structure  $\widehat{\mathfrak{bms}}_{4}\cong(\mathfrak{vir}\oplus\mathfrak{vir})\inplus\mathfrak{T}_{ab}$ where $(\mathfrak{vir}\oplus\mathfrak{vir})$ is spanned by  $\mathcal{L}$ and $\bar{\mathcal{L}}$ generators plus two {unit elements as} central generators and $\mathfrak{T}_{ab}$ is ideal part which is spanned by $T$.
The short exact sequence for the above is  
\begin{equation}
    0\longrightarrow \mathfrak{T}\longrightarrow \widehat{\mathfrak{bms}}_{4} \longrightarrow \widehat{\mathfrak{bms}}_{4}/\mathfrak{T}\cong (\mathfrak{vir}\oplus\mathfrak{vir})\longrightarrow 0.\label{short-exact-vir-vir}
\end{equation}

\paragraph{Computation of $\mathcal{H}^{2}(\widehat{\mathfrak{bms}}_{4};(\mathfrak{vir}\oplus\mathfrak{vir}))$.}
Using the Theorem 1.2 in \cite{degrijse2009cohomology} and Hochschild-Serre spectral sequence theorem we get
\begin{equation}
\begin{split}
     \hspace*{-5mm}\mathcal{H}^{2}(\widehat{\mathfrak{bms}}_{4};(\mathfrak{v}\oplus\mathfrak{v}))&=\oplus_{p+q=2}E_{2;(\mathfrak{v}\oplus\mathfrak{v})}^{p,q}=E_{2;(\mathfrak{v}\oplus\mathfrak{v})}^{2,0}\oplus E_{2;(\mathfrak{v}\oplus\mathfrak{v})}^{1,1}\oplus E_{2;(\mathfrak{v}\oplus\mathfrak{v})}^{0,2}\\
     &= \mathcal{H}^{2}((\mathfrak{vir}\oplus\mathfrak{vir});\mathcal{H}^{0}(\mathfrak{T},(\mathfrak{vir}\oplus\mathfrak{vir})))\cr &\oplus \mathcal{H}^{1}((\mathfrak{vir}\oplus\mathfrak{vir});\mathcal{H}^{1}(\mathfrak{T},(\mathfrak{vir}\oplus\mathfrak{vir})))\cr &\oplus\mathcal{H}^{0}((\mathfrak{vir}\oplus\mathfrak{vir});\mathcal{H}^{2}(\mathfrak{T},(\mathfrak{vir}\oplus\mathfrak{vir}))),
\end{split}
\end{equation}
where $E_{2;(\mathfrak{v}\oplus\mathfrak{v})}^{p,q}\equiv\mathcal{H}^{p}((\mathfrak{vir}\oplus\mathfrak{vir});\mathcal{H}^{q}(\mathfrak{T},(\mathfrak{vir}\oplus\mathfrak{vir})))$.

The first term we have to consider is $E_{2;(\mathfrak{v}\oplus\mathfrak{v})}^{2,0}=\mathcal{H}^{2}((\mathfrak{vir}\oplus\mathfrak{vir});\mathcal{H}^{0}(\mathfrak{T},(\mathfrak{vir}\oplus\mathfrak{vir})))$. From the definition of $\mathcal{H}^{0}$ one gets $\mathcal{H}^{0}(\mathfrak{T},(\mathfrak{vir}\oplus\mathfrak{vir}))=(\mathfrak{vir}\oplus\mathfrak{vir})$ because the action of $\mathfrak{T}$ as an ideal part of the algebra, on $(\mathfrak{vir}\oplus\mathfrak{vir})$, {induced by the short exact sequence \eqref{short-exact-vir-vir},} is trivial. Then, recalling the fact that $(\mathfrak{vir}\oplus\mathfrak{vir})$ algebra is rigid \cite{Parsa:2018kys} one concludes that $E_{2;(\mathfrak{vir}\oplus\mathfrak{vir})}^{2,0}=\mathcal{H}^{2}((\mathfrak{vir}\oplus\mathfrak{vir});(\mathfrak{vir}\oplus\mathfrak{vir}))=0$.

We should next consider $E_{2;(\mathfrak{v}\oplus\mathfrak{v})}^{1,1}=\mathcal{H}^{1}((\mathfrak{vir}\oplus\mathfrak{vir});\mathcal{H}^{1}(\mathfrak{T},(\mathfrak{vir}\oplus\mathfrak{vir})))$. One may generalize discussions of the case without the central elements to conclude
$\mathcal{H}^{1}((\mathfrak{vir}\oplus\mathfrak{vir});\mathcal{H}^{1}(\mathfrak{T},(\mathfrak{vir}\oplus\mathfrak{vir})))=0$. Therefore,  the $[\mathcal{L},T]$ and $[\bar{\mathcal{L}},T]$ commutators cannot be deformed by the terms with coefficient of $\hat{\mathcal{L}}$ and $\hat{\bar{\mathcal{L}}}$ where the hatted objects, such as $\hat{\mathcal{L}}_{m}$, denote generators of the Virasoro algebra, i.e. Witt algebra plus central element hence $E_{2;(\mathfrak{v}\oplus\mathfrak{v})}^{1,1}=0$.

The last term we compute is $E_{2;(\mathfrak{v}\oplus\mathfrak{v})}^{0,2}=\mathcal{H}^{0}((\mathfrak{vir}\oplus\mathfrak{vir});\mathcal{H}^{2}(\mathfrak{T},(\mathfrak{vir}\oplus\mathfrak{vir})))$. Using  definition of $\mathcal{H}^{0}$, one observes that its elements are solutions to
\begin{equation}
  (\hat{\mathcal{L}}_{m}\circ \hat{\psi})(T_{p,q},T_{r,s})=[\hat{\mathcal{L}}_{m},\hat{\psi}(T_{p,q},T_{r,s})]-\hat{\psi}([\hat{\mathcal{L}}_{m},T_{p,q}],T_{r,s})-\hat{\psi}(T_{p,q},[\hat{\mathcal{L}}_{m},T_{r,s}]),\label{psi-hat}
\end{equation}         
and
\begin{equation}
  (\bar{\hat{\mathcal{L}}}_{m}\circ \hat{\psi})(T_{p,q},T_{r,s})=[\bar{\hat{\mathcal{L}}}_{m},\hat{\psi}(T_{p,q},T_{r,s})]-\hat{\psi}([\bar{\hat{\mathcal{L}}}_{m},T_{p,q}],T_{r,s})-\hat{\psi}(T_{p,q},[\bar{\hat{\mathcal{L}}}_{m},T_{r,s}]),\label{psi-hat1}
\end{equation}    
where $\hat{\psi}(T_{p,q},T_{r,s})$ is a $2-$cocycle. 
The linear expansion of $\hat{\psi}$ in terms of generators is \[\hat{\psi}(T_{m,n},T_{p,q})=\tilde{A}(m,n;p,q)\mathcal{L}_{m+p-1}+\tilde{B}(m,n;p,q)\bar{\mathcal{L}}_{n+q-1}+\tilde{Z}(m,n;p,q).\] 
Inserting the expansion of $\hat{\psi}$ into \eqref{psi-hat} and \eqref{psi-hat1} we reach to some relations which force all of the above functions to be zero so $E_{2;(\mathfrak{v}\oplus\mathfrak{v})}^{0,2}=\mathcal{H}^{0}((\mathfrak{vir}\oplus\mathfrak{vir});\mathcal{H}^{2}(\mathfrak{T},(\mathfrak{vir}\oplus\mathfrak{vir})))=0$.

To conclude this part, we have shown that $\mathcal{H}^{2}(\widehat{\mathfrak{bms}}_{4};(\mathfrak{vir}\oplus\mathfrak{vir}))=0$. 

\paragraph{Computation of $\mathcal{H}^{2}(\widehat{\mathfrak{bms}}_{4};\mathfrak{T})$.}

Using the Theorem 1.2 in \cite{degrijse2009cohomology} and Hochschild-Serre spectral sequence theorem we get
\begin{equation}
\begin{split}
     \mathcal{H}^{2}(\widehat{\mathfrak{bms}}_{4};\mathfrak{T})&=\oplus_{p+q=2}E_{2;\mathfrak{T}}^{p,q}=E_{2;\mathfrak{T}}^{2,0}\oplus E_{2;\mathfrak{T}}^{1,1}\oplus E_{2;\mathfrak{T}}^{0,2}\cr
     &= \mathcal{H}^{2}((\mathfrak{vir}\oplus\mathfrak{vir});\mathcal{H}^{0}(\mathfrak{T},\mathfrak{T}))\oplus \mathcal{H}^{1}((\mathfrak{vir}\oplus\mathfrak{vir});\mathcal{H}^{1}(\mathfrak{T},\mathfrak{T}))\oplus\mathcal{H}^{0}((\mathfrak{vir}\oplus\mathfrak{vir});\mathcal{H}^{2}(\mathfrak{T},\mathfrak{T})),\nonumber
\end{split}
\end{equation}
where we defined $E_{2;\mathfrak{T}}^{p,q}=\mathcal{H}^{p}((\mathfrak{vir}\oplus\mathfrak{vir});\mathcal{H}^{q}(\mathfrak{T},\mathfrak{T}))$. Unlike the $\widehat{\mathfrak{bms}}_{3}$ case, the ideal part of the $\widehat{\mathfrak{bms}}_{4}$ do not admit any central generator in its ideal part, so the results of the case without central extension can be generalized to this case too. 

In summary, we conclude that 
\be\mathcal{H}^{2}(\widehat{\mathfrak{bms}}_{4},\widehat{\mathfrak{bms}}_{4})=\mathcal{H}^{1}((\mathfrak{vir}\oplus\mathfrak{vir});\mathcal{H}^{1}(\mathfrak{T};\mathfrak{T})).\label{bms-vir}
\end{equation}
That is, in accord with our closure condition analysis, $\widehat{\mathfrak{bms}}_4$ algebra can only be deformed to $\widehat{{\cal W}}(a,b;\bar{a},\bar{b})$.

\subsection{Cohomological consideration of \texorpdfstring{$\mathcal{W}(a,b;\bar{a},\bar{b})$}{W}  algebra} 


$\mathcal{W}(a,b;\bar{a},\bar{b})$ which is introduces by \eqref{W4-algebra} can be considered as semi-direct sum of $(\mathfrak{witt}\oplus\mathfrak{witt})\inplus \mathfrak{T}$ which it has the following short exact sequence 
\begin{equation}
     0\longrightarrow \mathfrak{T}\longrightarrow \mathcal{W}(a,b;\bar{a},\bar{b})\longrightarrow \mathcal{W}(a,b;\bar{a},\bar{b})/\mathfrak{T}\cong (\mathfrak{witt}\oplus\mathfrak{witt})\longrightarrow 0,
\end{equation}   
where $\mathfrak{T}$  and $(\mathfrak{witt}\oplus\mathfrak{witt})$, respectively, denote the ideal part and subalgebra of $\mathcal{W}(a,b;\bar{a},\bar{b})$. {Note that since $(\mathfrak{witt}\oplus\mathfrak{witt})$ is not a $\mathcal{W}(a,b;\bar{a},\bar{b})$ module by the adjoint action, $\mathcal{H}^2(\mathcal{W}(a,b;\bar{a},\bar{b});(\mathfrak{witt}\oplus\mathfrak{witt}))$ is defined by the action induced from the above short exact sequence. The relevant second adjoint cohomology of  $\mathcal{H}^2(\mathcal{W}(a,b;\bar{a},\bar{b});\mathcal{W}(a,b;\bar{a},\bar{b}))$ may hence be computed much the same as the \bmsf\ case discussed earlier, with the same result, namely,
$\mathcal{H}^2(\mathcal{W}(a,b;\bar{a},\bar{b}); \mathcal{W}(a,b;\bar{a},\bar{b}))=0$ for the \emph{family of \wfab\ algebras}; i.e. \wfab\ family  for generic value of parameters is infinitesimally and formally rigid.

\paragraph{Cohomological consideration of \texorpdfstring{$\widehat{\mathcal{W}}(a,b;\bar{a},\bar{b})$}{W}  algebra.}

The case $\widehat{\mathcal{W}}(a,b;\bar{a},\bar{b})$ is exactly the same as \hbmsf\ from cohomological point of view: It just admits two nontrivial central terms in its two Witt subalgebras and that $\mathcal{H}^2(\widehat{\mathcal{W}}(a,b;\bar{a},\bar{b}); \widehat{\mathcal{W}}(a,b;\bar{a},\bar{b}))=0$ for the family of $\widehat{\mathcal{W}}$ algebras.










\section{Summary and concluding remarks}\label{sec:8}

We analyzed the deformation and stability (rigidity) of \bmsf\ and its central extension \hbmsf\ algebra  which appear as the asymptotic symmetry of 4d flat spacetime. We showed that although they are stable (rigid) in their ideal part, 
they can be formally deformed in other commutators which lead to a family of new non-isomorphic infinite dimensional algebras we called $\mathcal{W}(a,b;\bar{a},\bar{b})$ and $\widehat{\mathcal{W}}(a,b;\bar{a},\bar{b})$. Among other things, this implies that there is no such infinite dimensional algebra associated with AdS$_4$ space which upon contraction yields \bmsf\ or \hbmsf. This is unlike the 3d case where \hbmst\ and $\mathfrak{vir}\oplus\mathfrak{vir}$ are related through deformation/contraction, which prompts  a possible formulation of field theory holographically dual to gravity on 3d flat space \cite{Barnich:2010eb, Bagchi:2010eg, Hartong:2015usd, Bagchi:2016bcd}. In the 4d case one should seek a different way to tackle the question of field theories dual to gravity on 4d flat space.

We also showed that while the family of $\mathcal{W}(a,b;\bar{a},\bar{b})$ and $\widehat{\mathcal{W}}(a,b;\bar{a},\bar{b})$ are stable (rigid) for generic values of their parameters,  for specific points  such as $\widehat{\mathcal{W}}(0,0;0,0)$ and $\widehat{\mathcal{W}}(0,-1;0,0)$, there are other possibilities for deformation which takes us out of the ${\cal W}$-algebra family. 
As the first interesting follow-up question one may explore whether $\mathcal{W}(a,b;\bar{a},\bar{b})$ and $\widehat{\mathcal{W}}(a,b;\bar{a},\bar{b})$ algebras may appear as asymptotic symmetries of a physical theory. One may first tackle a similar question in the simpler case of $W(a,b)$ algebra discussed in \cite{gao2011low, Parsa:2018kys}. The next natural question is then what is the physical/geometric meaning of the deformations and motion in the parameter space of the algebras.


\paragraph{Hochschild-Serre factorization (HSF) theorem for infinite dimensional algebras?!} We already discussed the HSF theorem states that a non-rigid \emph{finite dimensional} algebra can only be deformed in its ideal part and other parts of the algebra cannot be deformed. It is, however, known that this theorem does not apply to infinite dimensional algebras. Our analysis in \cite{Parsa:2018kys} for the \bmst,  $\mathfrak{KM}_{u(1)}$ and $W(a,b)$, led us to a proposal for a version of 
HSF which works for the infinite dimensional algebras: \emph{For infinite dimensional algebras with countable basis the deformations may appear in ideal and non-ideal parts, however, the deformations are always by coefficients of term in the ideal part.} For the $W(a,b)$ class with generators ${\cal L}_n, {\cal P}_n$ the deformations may appear in $[{\cal L}, {\cal L}],\ [{\cal L}, {\cal P}],\ [{\cal P}, {\cal P}],$ nonetheless it is proportional to ${\cal P}$ generators. Our analysis of \bmsf, \wfab\ algebras confirm this extended HSF theorem, providing more supportive examples for it. It would be desirable to attempt proving this proposal, maybe using cohomological arguments.



As we mentioned although the $\mathcal{W}(a,b;\bar{a},\bar{b})$ algebrafamily  and their central extensions are stable and cannot be deformed, there are some specific points which can be deformed in other commutators. The most interesting case is the $\mathcal{W}(0,-1;0,0)$ algebra which admits a formal deformation in its ideal part. The central extension of $\mathcal{W}(0,-1;0,0)$ algebra can also be deformed to a new algebra which has a direct sum of three Virasoro subalgebras. As we discussed the $\mathcal{W}(0,-1;0,0)$ algebra has $\mathfrak{iso}(2,1)\oplus\mathfrak{sl}(2,\mathbb{R})$ global part. By deformation of the $\mathcal{W}(0,-1;0,0)$ algebra we obtain a new algebra which has $\mathfrak{so}(2,2)\oplus\mathfrak{so}(2,1)$ ($\mathfrak{so}(3,1)\oplus\mathfrak{so}(2,1)$) which is direct sum of isometry algebra of AdS$_3$ (dS$_3$) spacetime with 3d Lorentz algebra \cite{gomis2009deformations, salgado2014so}.  It is interesting to explore if this subalgebra 
is related to the asymptotic symmetry algebra of 3d Einstein-Maxwell theory (see \cite{Barnich:2015jua, Concha:2018zeb,Caroca:2017onr}) or to 
the ``meta'' conformal algebras \cite{Henkel:2017enn} which can have a connection with conformal field theory.

Here we focused on the algebras and their deformations. We know that there are groups associated with \bmsf\ and its central extension \hbmsf\ algebras \cite{Barnich:2016lyg}. One may ask how the deformation of algebras appear in the associated groups, e.g. whether there are groups associated with $\mathcal{W}(a,b;\bar{a},\bar{b})$ and its central extension $\widehat{\mathcal{W}}(a,b;\bar{a},\bar{b})$ algebras. Another related question is analyzing the (co)adjoint orbits of these groups and algebras and how the deformations affect the coadjoint orbits
which is crucial for building Hilbert space of physical theories  invariant under the deformed symmetry algebras.

\section*{Acknowledgment}

We acknowledge partial support of Iranian NSF Junior Research Chair under grant No. 950124 and Saramadan-Iran grant no. ISEF/M/97219. We would like to thank Hamid Afshar, Glenn Barnich and Daniel Grumiller for comments on the draft. 


\appendix
\section{Algebra generators as functions on celestial two sphere}\label{appendix-A}

The \bmst\ and \bmsf\ algebras may be obtained as asymptotic symmetry algebra 3d and 4d flat spacetimes, respectively. As such the generators of these algebras are given through co-dimension two surface integrals which is an integral over a circle for \bmst\ case and over a two-sphere for the \bmsf\ case.
In other words, the generators of \bmst\ and \bmsf\ algebras respectively may be viewed as functions on an $S^1$ or $S^2$. In fact, it is known that $\mathfrak{witt}$ (or Virasoro algebra) is nothing but algebra of diffeomorphisms on an $S^1$. In this appendix we explore this viewpoint and its implications.

\paragraph{$\mathfrak{witt}$ algebra case,  a warm up example.} Let us first analyze the case of $\mathfrak{witt}$. The case of \bmst\ was analyzed in \cite{Parsa:2018kys}. Generators of the $\mathfrak{witt}$ are Fourier modes of a (periodic) function on $S^1$:
\begin{equation}
    \mathcal{L}_{n}=\frac{1}{2\pi}\int^{2\pi}_{0} d\varphi \mathcal{L}(\varphi)\exp{(in\varphi)}.
\end{equation}
The index $n$ on the generators, hence, depends on the Fourier basis used. Explicitly, one may use 
\begin{equation}\begin{split}
    \tilde{\mathcal{L}}_{n}&=\frac{1}{2\pi}\int^{2\pi}_{0}  d\varphi \mathcal{L}(\varphi)\exp{(in\varphi)}{\Phi'(\varphi)}\cr
    &= \frac{1}{2\pi}\int^{2\pi}_{0}  d\Phi   \tilde{\mathcal{L}}(\Phi)\exp{(in\varphi(\Phi))},
    \end{split}
\end{equation}
where  $\Phi(\varphi)$ is some periodic function, $\varphi(\Phi)$ is its inverse and $\tilde{\mathcal{L}}(\Phi)=\mathcal{L}(\varphi)$. For example, if we choose 
$\Phi=\frac{K}{d_{0}} e^{id_{0}\varphi}$, then  $\tilde{\mathcal{L}}_{n}=\sum_{d}C_{d}\mathcal{L}_{n+d}$ with $C_{d}=(\frac{K}{d_{0}})^{l}\frac{1}{l!}$ in which $K$ is a constant number. 

The above simple analysis shows that one have the freedom to shift the index $n$ on ${\cal L}_n$ generators an go to another equivalent basis; this change of basis geometrically corresponds to  a diffeomorphism on circle.

\paragraph{$\mathfrak{bms}_{4}$ algebra case.} We are now ready to make a similar analysis for \bmsf\ case where the generators are function on an $S^2$ \cite{Barnich:2009se, Barnich:2011ct}. To this end, let adopt Poincar\'e coordinates for the $S^2$,
$ds^{2}=\frac{1}{(1+\zeta\bar{\zeta})^{2}}d\zeta d\bar{\zeta}$. In this coordinates the standard basis for expansion of generators are
\begin{equation}
\begin{split}
{\cal L}(\zeta)=\sum {\cal L}_n \zeta^{n+1}&,\qquad \bar{\mathcal{L}}(\zeta)=\sum \bar{\mathcal{L}}_n \bar{\zeta}^{n+1},\cr
    T(\zeta,\bar{\zeta})&=\sum T_{m,n}\zeta^{n}\bar{\zeta}^{m}
\end{split}
\end{equation}
where $\zeta,\bar{\zeta}$ are coordinates on the sphere. A change of coordinates on the $S^2$, for example like the one discuss above for the case of a circle, yields a change in the indices on ${\cal L}_n, \bar{\mathcal{L}}_n, T_{m,n}$. Fixing this coordinates, fixes the conventions on the indices.\footnote{Of course, recalling that the global part of the supertranslations $T_{00},T_{01},T_{10}, T_{11}$ are in the $(2,2)$ representation of the Lorentz group $\mathfrak{su}(2)_L\times \mathfrak{su}(2)_R$, it is also natural to choose the indices to be half-integer valued, as suggested in \cite{Barnich:2017ubf}. } 
Analyzing the commutators and hence in what can appear on the Right-Hand-Side of the deformed commutators,  we are dealing with product of these functions. Let us e.g. consider $T^{(1)}(\zeta,\bar{\zeta})T^{(2)}(\zeta,\bar{\zeta})$,
\begin{equation}
    T^{(1)}(\zeta,\bar{\zeta})T^{(2)}(\zeta,\bar{\zeta})=\sum T^{(1)}_{m,n}T^{(2)}_{p,q}\zeta^{m+p}\bar{\zeta}^{n+q}\equiv \sum T_{k,l}\zeta^{k}\bar{\zeta}^{l}\ \Longrightarrow\ 
    T_{k,l}=\sum T^{(1)}_{m,n}T^{(2)}_{k-m,l-n},
\end{equation}
so one finds that the index $k,l$ are fixed to be sum indices of $T^{(1)}_{m,n}$ and $T^{(2)}_{k-m,l-n}$. In a similar way, the indices in deformations of $[{\cal L}_n,{\cal L}_m], [{\cal L}_n, \bar{\mathcal{L}}_m], [{\cal L}_n, T_{p,q}], [{\cal L}_n, \bar{\mathcal{L}}_m], [\bar{\mathcal{L}}_m,\bar{\mathcal{L}}_n]$ and $[\bar{\mathcal{L}}_m, T_{p,q}]$ are fixed.

\section{Hochschild-Serre spectral sequence }\label{appendix-B}

It is known that  Hochschild-Serre factorization theorem does not work infinite dimensional Lie algebras, nonetheless the Hochschild-Serre spectral sequence method applies also to infinite dimensional cases and can be used to extract information about deformations and hence second adjoint cohomology.  
For a Lie algebra $(\mathfrak{g},[,])$ {with a semi-direct sum structure as $\mathfrak{g}=\mathfrak{g}_{0}\inplus \mathfrak{h}$ where $\mathfrak{h}$ is an abelian ideal and $\mathfrak{g}_{0}\cong \mathfrak{g}/\mathfrak{h}$ is its quotient Lie algebra}, we have the following short exact sequence
\begin{equation}
    0\longrightarrow \mathfrak{h}\longrightarrow \mathfrak{g} \longrightarrow \mathfrak{g}/\mathfrak{h}\cong \mathfrak{g}_{0}\longrightarrow 0,\label{short-exact}
\end{equation}
where arrows show {Lie algebra morphisms. The arrows in the short exact sequence means the image of each morphism} is equal to the kernel of the next. For this sequence one obtains the Hochschild-Serre spectral sequence of cochain complexes whose  first terms are (see \cite{Parsa:2018kys} for a more detailed review and references)
    \begin{equation*} 
\begin{split}\label{chain-complex}
 E_{0}^{p,q}=\mathcal{C}^{p}(\mathfrak{g}_{0},\mathcal{C}^{q}(\mathfrak{h},M)),\ E_{1}^{p,q}=\mathcal{H}^{p}(\mathfrak{g}_{0};\mathcal{C}^{q}(\mathfrak{h},M)),\ 
   E_{2}^{p,q}=\mathcal{H}^{p}(\mathfrak{g}_{0};\mathcal{H}^{q}(\mathfrak{h};M)),\ ...,\ E_{n}^{p,q},... 
\end{split}
\end{equation*}
in which $M$ is a $\mathfrak{g}$-module, $\mathcal{C}^{p}$  is the space of $p$-cochains and $E$'s are related to each other by the differential operator $d_{n}^{p,q}:E_{n}^{p,q}\longrightarrow E_{n}^{p+n,q-n+1}$ \cite{H-S-factorization-theorem,fuks2012cohomology}. In some specific cases one finds  the differential function becomes trivial for $n\geq n_0$ (for certain $n_0$) and $E_{n}^{p,q}, \forall n\geq n_0$ are isomorphic to each other and therefore, $E_{n}^{p,q}\cong E_{\infty}^{p,q}$. So for the latter we have \footnote{Note that,  in general, this equality is true modulo extensions but all the terms in our cases are vector spaces and hence those extensions do not appear.}
\begin{equation}
    \mathcal{H}^{n}(\mathfrak{g};M)=\oplus_{p+q=n}E_{\infty}^{p,q}.\label{dec}
\end{equation}
{In this setting by the Hochschild-Serre spectral theorem \cite{H-S-factorization-theorem} we have} 
\begin{equation}
E_{2}^{p,q}=
    \mathcal{H}^{p}(\mathfrak{g}_{0}; \mathcal{H}^{q}(\mathfrak{h},M)).\label{E2-dec}
\end{equation}
 This theorem works for both finite and  infinite  dimensional Lie algebras.
{For those split abelian extensions with the property that the ideal action on $M$ is trivial, Theorem 1.2 in \cite{degrijse2009cohomology} states that 
we always have} $n_{0}=2$ and {therefore} $E_{2}^{p,q}\cong E_{\infty}^{p,q}$. So, combining \eqref{dec} and \eqref{E2-dec} one finds 
\begin{equation}
    \mathcal{H}^{2}(\mathfrak{g};M)=\oplus_{p+q=2}E_{2}^{p,q}.\label{dec2}
\end{equation}

{Note that $\mathfrak{h}$ is an ideal of $\mathfrak{g}$ and hence a $\mathfrak{g}$-module and because it is abelian, as a $\mathfrak{g}$-module its action on itself is trivial. Using the short exact sequence (\ref{short-exact}) we consider 
 $\mathfrak{g}_{0}$ as a $\mathfrak{g}$-module as well. In this way the action of $\mathfrak{h}$ on $\mathfrak{g}_{0}$ is trivial. We conclude that via the above arguments, $\mathfrak{g}_{0}$ and $\mathfrak{h}$ are both $\mathfrak{g}$-modules satisfying the conditions of Theorem 1.2 in \cite{degrijse2009cohomology}, and one can compute the spaces $\mathcal{H}^{2}(\mathfrak{g};\mathfrak{g}_{0})$ and $\mathcal{H}^{2}(\mathfrak{g};\mathfrak{h})$.} 
 
The short exact sequence \eqref{short-exact} induces the long exact sequence at the level of cohomologies 
\begin{equation}
    \begin{split} & \cdots \longrightarrow \mathcal{H}^{1}(\mathfrak{g};\mathfrak{g}_{0})\longrightarrow \mathcal{H}^{2}(\mathfrak{g};\mathfrak{h}) \longrightarrow \mathcal{H}^{2}(\mathfrak{g};\mathfrak{g})\longrightarrow \mathcal{H}^{2}(\mathfrak{g};\mathfrak{g}_{0})
     \longrightarrow \mathcal{H}^{3}(\mathfrak{g};\mathfrak{g}_{0}) \longrightarrow \cdots \label{long-exact}
    \end{split}
\end{equation}
One may use the above sequence to get information about  $\mathcal{H}^{2}(\mathfrak{g};\mathfrak{g})$ or even compute it. The long exact sequence \eqref{long-exact} is true for both finite and infinite dimensional Lie algebras with the semi-direct sum structure. In finite dimensional cases as a consequence of Hochschild-Serre factorization theorem we have $ \mathcal{H}^{2}(\mathfrak{g};\mathfrak{g})\cong \mathcal{H}^{2}(\mathfrak{g};\mathfrak{h})$. The latter can be obtained from another long exact sequence 
 \begin{equation}
    \begin{split} & \cdots \longrightarrow \mathcal{H}^{1}(\mathfrak{h};\mathfrak{g})\longrightarrow \mathcal{H}^{2}(\mathfrak{g}_{0};\mathfrak{g}) \longrightarrow \mathcal{H}^{2}(\mathfrak{g};\mathfrak{g})\longrightarrow \mathcal{H}^{2}(\mathfrak{h};\mathfrak{g})
     \longrightarrow \mathcal{H}^{3}(\mathfrak{g}_{0};\mathfrak{g}) \longrightarrow \cdots \label{long-exact2}
    \end{split}
\end{equation}
in which  $\mathcal{H}^{2}(\mathfrak{g}_{0};\mathfrak{g})=\mathcal{H}^{3}(\mathfrak{g}_{0};\mathfrak{g})=0$.
 In the case of infinite dimensional Lie algebras we can still use \eqref{long-exact}. While the sequence has in general infinite terms, in some specific cases one finds that some of terms in \eqref{long-exact} are equal to zero, leading to another  short exact sequence. In such situations we can infer some information about lower cohomologies, see \cite{Parsa:2018kys} for some examples.
\bibliographystyle{fullsort.bst}
  \providecommand{\href}[2]{#2}\begingroup\raggedright\endgroup


\begin{thebibliography}{10}
 	
 	\bibitem{Parsa:2018kys}
 	A.~Farahmand~Parsa, H.~R. Safari, and M.~M. Sheikh-Jabbari, ``{On Rigidity of
 		3d Asymptotic Symmetry Algebras},''
 	\href{http://www.arXiv.org/abs/1809.08209}{{\tt 1809.08209}}.
 	
 	\bibitem{Hawking:2016msc}
 	S.~W. Hawking, M.~J. Perry, and A.~Strominger, ``{Soft Hair on Black Holes},''
 	{\em Phys. Rev. Lett.} {\bf 116} (2016), no.~23, 231301,
 	\href{http://www.arXiv.org/abs/1601.00921}{{\tt 1601.00921}}.
 	
 	\bibitem{Hawking:2016sgy}
 	S.~W. Hawking, M.~J. Perry, and A.~Strominger, ``{Superrotation Charge and
 		Supertranslation Hair on Black Holes},'' {\em JHEP} {\bf 05} (2017) 161,
 	\href{http://www.arXiv.org/abs/1611.09175}{{\tt 1611.09175}}.
 	
 	\bibitem{Strominger:2017zoo}
 	A.~Strominger, ``{Lectures on the Infrared Structure of Gravity and Gauge
 		Theory},''
 	\href{http://www.arXiv.org/abs/1703.05448}{{\tt 1703.05448}}.
 	
 	\bibitem{Ashtekar:1996cd}
 	{Ashtekar, Abhay and Bi{\v{c}}{\'a}k, Ji{\v{r}}{\'\i} and Schmidt, Bernd G},
 	``{Asymptotic structure of symmetry reduced general relativity},'' {\em Phys.
 		Rev.} {\bf D55} (1997) 669--686,
 	\href{http://www.arXiv.org/abs/gr-qc/9608042}{{\tt gr-qc/9608042}}.
 	
 	\bibitem{Oblak:2016eij}
 	B.~Oblak, {\em {BMS Particles in Three Dimensions}}.
 	\newblock PhD thesis, Brussels U., 2016.
 	\newblock
 	\href{http://www.arXiv.org/abs/1610.08526}{{\tt 1610.08526}}.
 	\newblock
 	
 	\bibitem{Bondi:1962px}
 	H.~Bondi, M.~G.~J. van~der Burg, and A.~W.~K. Metzner, ``{Gravitational waves
 		in general relativity. 7. Waves from axisymmetric isolated systems},'' {\em
 		Proc. Roy. Soc. Lond.} {\bf A269} (1962)
 	21--52.
 	
 	\bibitem{Sachs:1962zza}
 	R.~Sachs, ``{Asymptotic symmetries in gravitational theory},'' {\em Phys. Rev.}
 	{\bf 128} (1962) 2851--2864.
 	
 	\bibitem{Sachs:1962wk}
 	R.~K. Sachs, ``{Gravitational waves in general relativity. 8. Waves in
 		asymptotically flat space-times},'' {\em Proc. Roy. Soc. Lond.} {\bf A270}
 	(1962) 103--126.
 	
 	\bibitem{Barnich:2006av}
 	G.~Barnich and G.~Compere, ``{Classical central extension for asymptotic
 		symmetries at null infinity in three spacetime dimensions},'' {\em Class.
 		Quant. Grav.} {\bf 24} (2007) F15--F23,
 	\href{http://www.arXiv.org/abs/gr-qc/0610130}{{\tt gr-qc/0610130}}.
 	
 	\bibitem{Barnich:2009se}
 	G.~Barnich and C.~Troessaert, ``{Symmetries of asymptotically flat 4
 		dimensional spacetimes at null infinity revisited},'' {\em Phys. Rev. Lett.}
 	{\bf 105} (2010) 111103,
 	\href{http://www.arXiv.org/abs/0909.2617}{{\tt 0909.2617}}.
 	
 	\bibitem{Barnich:2011mi}
 	G.~Barnich and C.~Troessaert, ``{BMS charge algebra},'' {\em JHEP} {\bf 12}
 	(2011) 105,
 	\href{http://www.arXiv.org/abs/1106.0213}{{\tt 1106.0213}}.
 	
 	\bibitem{Duval:2014uva}
 	C.~Duval, G.~W. Gibbons, and P.~A. Horvathy, ``{Conformal Carroll groups and
 		BMS symmetry},'' {\em Class. Quant. Grav.} {\bf 31} (2014) 092001,
 	\href{http://www.arXiv.org/abs/1402.5894}{{\tt 1402.5894}}.
 	
 	\bibitem{Barnich:2017ubf}
 	G.~Barnich, ``{Centrally extended BMS4 Lie algebroid},'' {\em JHEP} {\bf 06}
 	(2017) 007,
 	\href{http://www.arXiv.org/abs/1703.08704}{{\tt 1703.08704}}.
 	
 	\bibitem{troessaert2018bms4}
 	C.~Troessaert, ``The $\text{BMS}_{4}$ algebra at spatial infinity,'' {\em
 		Classical and Quantum Gravity} {\bf 35} (2018), no.~7, 074003.
 	
 	\bibitem{Brown:1986nw}
 	J.~D. Brown and M.~Henneaux, ``{Central Charges in the Canonical Realization of
 		Asymptotic Symmetries: An Example from Three-Dimensional Gravity},'' {\em
 		Commun. Math. Phys.} {\bf 104} (1986)
 	207--226.
 	
 	\bibitem{Ashtekar:1984zz}
 	A.~Ashtekar and A.~Magnon, ``{Asymptotically anti-de Sitter space-times},''
 	{\em Class. Quant. Grav.} {\bf 1} (1984)
 	L39--L44.
 	
 	\bibitem{Compere:2013bya}
 	G.~Comp\`ere, W.~Song, and A.~Strominger, ``{New Boundary Conditions for
 		$\text{AdS}_3$},'' {\em JHEP} {\bf 05} (2013) 152,
 	\href{http://www.arXiv.org/abs/1303.2662}{{\tt 1303.2662}}.
 	
 	\bibitem{Afshar:2016wfy}
 	H.~Afshar, S.~Detournay, D.~Grumiller, W.~Merbis, A.~Perez, D.~Tempo, and
 	R.~Troncoso, ``{Soft Heisenberg hair on black holes in three dimensions},''
 	{\em Phys. Rev.} {\bf D93} (2016), no.~10, 101503,
 	\href{http://www.arXiv.org/abs/1603.04824}{{\tt 1603.04824}}.
 	
 	\bibitem{Afshar:2016kjj}
 	H.~Afshar, D.~Grumiller, W.~Merbis, A.~Perez, D.~Tempo, and R.~Troncoso,
 	``{Soft hairy horizons in three spacetime dimensions},'' {\em Phys. Rev.}
 	{\bf D95} (2017), no.~10, 106005,
 	\href{http://www.arXiv.org/abs/1611.09783}{{\tt 1611.09783}}.
 	
 	\bibitem{Afshar:2017okz}
 	H.~Afshar, D.~Grumiller, M.~M. Sheikh-Jabbari, and H.~Yavartanoo, ``{Horizon
 		fluff, semi-classical black hole microstates — Log-corrections to BTZ
 		entropy and black hole/particle correspondence},'' {\em JHEP} {\bf 08} (2017)
 	087,
 	\href{http://www.arXiv.org/abs/1705.06257}{{\tt 1705.06257}}.
 	
 	\bibitem{Grumiller:2016pqb}
 	D.~Grumiller and M.~Riegler, ``{Most general AdS$_{3}$ boundary conditions},''
 	{\em JHEP} {\bf 10} (2016) 023,
 	\href{http://www.arXiv.org/abs/1608.01308}{{\tt 1608.01308}}.
 	
 	\bibitem{Grumiller:2017sjh}
 	D.~Grumiller, W.~Merbis, and M.~Riegler, ``{Most general flat space boundary
 		conditions in three-dimensional Einstein gravity},'' {\em Class. Quant.
 		Grav.} {\bf 34} (2017), no.~18, 184001,
 	\href{http://www.arXiv.org/abs/1704.07419}{{\tt 1704.07419}}.
 	
 	\bibitem{donnay:2015abr}
 	L.~Donnay, G.~Giribet, H.~A. Gonzalez, and M.~Pino, ``{Supertranslations and
 		Superrotations at the Black Hole Horizon},'' {\em Phys. Rev. Lett.} {\bf 116}
 	(2016), no.~9, 091101,
 	\href{http://www.arXiv.org/abs/1511.08687}{{\tt 1511.08687}}.
 	
 	\bibitem{NH-symmetry}
 	D.~Grumiller, A.~Perez, M.~Sheikh-Jabbari, R.~Troncoso, and C.~Zwikel, ``{Soft
 		hair on black hole and cosmological horizons in any dimension}.'' {\it To
 		appear}.
 	
 	\bibitem{grumiller:2018scv}
 	D.~Grumiller and M.~M. Sheikh-Jabbari, ``{Membrane Paradigm from Near Horizon
 		Soft Hair},'' {\em Int. J. Mod. Phys.} {\bf D27} (2018), no.~14, 1847006,
 	\href{http://www.arXiv.org/abs/1805.11099}{{\tt 1805.11099}}.
 	
 	\bibitem{kapec2014asymptotic}
 	D.~Kapec, V.~Lysov, and A.~Strominger, ``Asymptotic symmetries of massless
 	$\text{QED}$ in even dimensions,'' {\em arXiv preprint arXiv:1412.2763}
 	(2014).
 	
 	\bibitem{Hosseinzadeh:2018dkh}
 	V.~Hosseinzadeh, A.~Seraj, and M.~M. Sheikh-Jabbari, ``{Soft Charges and
 		Electric-Magnetic Duality},'' {\em JHEP} {\bf 08} (2018) 102,
 	\href{http://www.arXiv.org/abs/1806.01901}{{\tt 1806.01901}}.
 	
 	\bibitem{strominger:2015bla}
 	A.~Strominger, ``{Magnetic Corrections to the Soft Photon Theorem},'' {\em
 		Phys. Rev. Lett.} {\bf 116} (2016), no.~3, 031602,
 	\href{http://www.arXiv.org/abs/1509.00543}{{\tt 1509.00543}}.
 	
 	\bibitem{campiglia:2016hvg}
 	M.~Campiglia and A.~Laddha, ``{Subleading soft photons and large gauge
 		transformations},'' {\em JHEP} {\bf 11} (2016) 012,
 	\href{http://www.arXiv.org/abs/1605.09677}{{\tt 1605.09677}}.
 	
 	\bibitem{Afshar:2018apx}
 	H.~Afshar, E.~Esmaeili, and M.~M. Sheikh-Jabbari, ``{Asymptotic Symmetries in
 		$p$-Form Theories},'' {\em JHEP} {\bf 05} (2018) 042,
 	\href{http://www.arXiv.org/abs/1801.07752}{{\tt 1801.07752}}.
 	
 	\bibitem{Francia:2018jtb}
 	D.~Francia and C.~Heissenberg, ``{Two-Form Asymptotic Symmetries and Scalar
 		Soft Theorems},'' {\em Phys. Rev.} {\bf D98} (2018), no.~10, 105003,
 	\href{http://www.arXiv.org/abs/1810.05634}{{\tt 1810.05634}}.
 	
 	\bibitem{gerstenhaber1964deformation}
 	M.~Gerstenhaber, ``On the deformation of rings and algebras: I,'' {\em Ann. Of
 		Math} (1964) 59--103.
 	
 	\bibitem{gerstenhaber1966deformation}
 	M.~Gerstenhaber, ``On the deformation of rings and algebras: {II},'' {\em Ann.
 		Of Math} (1966) 1--19.
 	
 	\bibitem{gerstenhaber1968deformation}
 	M.~Gerstenhaber, ``On the deformation of rings and algebras: {III},'' {\em Ann.
 		Of Math} (1968) 1--34.
 	
 	\bibitem{gerstenhaber1974deformation}
 	M.~Gerstenhaber, ``On the deformation of rings and algebras: {IV},'' {\em Ann.
 		Of Math.} (1974) 257--276.
 	
 	\bibitem{nijenhuis1967deformations}
 	A.~Nijenhuis and R.~Richardson, ``Deformations of lie algebra structures,''
 	{\em Journal of Mathematics and Mechanics} {\bf 17} (1967), no.~1, 89--105.
 	
 	\bibitem{levy1967deformation}
 	M.~Levy-Nahas, ``Deformation and contraction of {L}ie algebras,'' {\em J. Math.
 		Phys.} {\bf 8} (1967), no.~6, 1211--1222.
 	
 	\bibitem{Whitehead-1}
 	J.~Whitehead, ``Combinatorial homotopy. {I},'' {\em Bull. Amer. Math. Soc.}
 	{\bf 55} (1949) 213--245.
 	
 	\bibitem{Whitehead-2}
 	J.~Whitehead, ``Combinatorial homotopy. {II},'' {\em Bull. Amer. Math. Soc.}
 	{\bf 55} (1949) 453--496.
 	
 	\bibitem{H-S-factorization-theorem}
 	G.~Hochschild and J.-P. Serre, ``Cohomology of lie algebras,'' {\em Annals of
 		Mathematics} {\bf 57} (1953), no.~3, 591--603.
 	
 	\bibitem{Inonu:1953sp}
 	E.~In{\"o}n{\"u} and E.~Wigner, ``{On the contraction of groups and their
 		representations},'' {\em Proc. Nat. Acad. Sci.} {\bf 39} (1953) 510--524.
 	
 	\bibitem{mendes1994deformations}
 	R.~V. Mendes, ``Deformations, stable theories and fundamental constants,'' {\em
 		J. Phys. A.} {\bf 27} (1994), no.~24, 8091.
 	
 	\bibitem{Figueroa-OFarrill:1989wmj}
 	J.~M. Figueroa-O'Farrill, ``{Deformations of the Galilean Algebra},'' {\em J.
 		Math. Phys.} {\bf 30} (1989)
 	2735.
 	
 	\bibitem{Chryssomalakos:2004gk}
 	C.~Chryssomalakos and E.~Okon, ``{Generalized quantum relativistic kinematics:
 		A Stability point of view},'' {\em Int. J. Mod. Phys.} {\bf D13} (2004)
 	2003--2034,
 	\href{http://www.arXiv.org/abs/hep-th/0410212}{{\tt hep-th/0410212}}.
 	
 	\bibitem{Figueroa-OFarrill:2017sfs}
 	J.~Figueroa-O'Farrill, ``{Classification of kinematical Lie algebras},''
 	\href{http://www.arXiv.org/abs/1711.05676}{{\tt 1711.05676}}.
 	
 	\bibitem{Figueroa-OFarrill:2017ycu}
 	J.~M. Figueroa-O'Farrill, ``{Kinematical Lie algebras via deformation
 		theory},'' {\em J. Math. Phys.} {\bf 59} (2018), no.~6, 061701,
 	\href{http://www.arXiv.org/abs/1711.06111}{{\tt 1711.06111}}.
 	
 	\bibitem{Figueroa-OFarrill:2017tcy}
 	J.~M. Figueroa-O'Farrill, ``{Higher-dimensional kinematical Lie algebras via
 		deformation theory},'' {\em J. Math. Phys.} {\bf 59} (2018), no.~6, 061702,
 	\href{http://www.arXiv.org/abs/1711.07363}{{\tt 1711.07363}}.
 	
 	\bibitem{Andrzejewski:2018gmz}
 	T.~Andrzejewski and J.~Figueroa-O'Farrill, ``{Kinematical Lie algebras in 2+1
 		dimensions},'' {\em J. Math. Phys.} {\bf 59} (2018), no.~6, 061703,
 	\href{http://www.arXiv.org/abs/1802.04048}{{\tt 1802.04048}}.
 	
 	\bibitem{Figueroa-OFarrill:2018ygf}
 	J.~M. Figueroa-O'Farrill, ``{Conformal Lie algebras via deformation theory},''
 	\href{http://www.arXiv.org/abs/1809.03603}{{\tt 1809.03603}}.
 	
 	\bibitem{Figueroa-OFarrill:2018ilb}
 	J.~Figueroa-O'Farrill and S.~Prohazka, ``{Spatially isotropic homogeneous
 		spacetimes},''
 	\href{http://www.arXiv.org/abs/1809.01224}{{\tt 1809.01224}}.
 	
 	\bibitem{Fialowski:2001me}
 	A.~Fialowski and M.~Penkava, ``{Deformation Theory of Infinity Algebras},''
 	{\em J. Algebra.} {\bf 255} (2002) 59--88,
 	\href{http://www.arXiv.org/abs/math/0101097}{{\tt math/0101097}}.
 	
 	\bibitem{fialowski2012formal}
 	A.~Fialowski, ``Formal rigidity of the witt and virasoro algebra,'' {\em J.
 		Mat. Phys.} {\bf 53} (2012), no.~7, 073501.
 	
 	\bibitem{gao2008derivations}
 	S.~Gao, C.~Jiang, and Y.~Pei, ``The derivations, central extensions and
 	automorphism group of the lie algebra \emph{W},''
 	\href{http://www.arXiv.org/abs/arXiv:0801.3911v1}{{\tt arXiv:0801.3911v1}}.
 	
 	\bibitem{gao2011low}
 	S.~Gao, C.~Jiang, and Y.~Pei, ``Low-dimensional cohomology groups of the lie
 	algebras $\text{W}(a, b)$,'' {\em Commun. Algebra} {\bf 39} (2011), no.~2,
 	397--423.
 	
 	\bibitem{Ecker:2017sen}
 	J.~Ecker and M.~Schlichenmaier, ``{The Vanishing of the Low-Dimensional
 		Cohomology of the Witt and the Virasoro algebra},''
 	\href{http://www.arXiv.org/abs/1707.06106}{{\tt 1707.06106}}.
 	
 	\bibitem{Ecker:2018iqn}
 	J.~Ecker and M.~Schlichenmaier, ``{The Low-Dimensional Algebraic Cohomology of
 		the Virasoro Algebra},''
 	\href{http://www.arXiv.org/abs/1805.08433}{{\tt 1805.08433}}.
 	
 	\bibitem{Barnich:2012rz}
 	G.~Barnich, A.~Gomberoff, and H.~A. Gonzalez, ``{Three-dimensional
 		Bondi-Metzner-Sachs invariant two-dimensional field theories as the flat
 		limit of Liouville theory},'' {\em Phys. Rev.} {\bf D87} (2013), no.~12,
 	124032,
 	\href{http://www.arXiv.org/abs/1210.0731}{{\tt 1210.0731}}.
 	
 	\bibitem{Barnich:2011ct}
 	G.~Barnich and C.~Troessaert, ``{Supertranslations call for superrotations},''
 	{\em PoS} {\bf CNCFG2010} (2010) 010,
 	\href{http://www.arXiv.org/abs/1102.4632}{{\tt 1102.4632}}.
 	[Ann. U. Craiova Phys.21,S11(2011)].
 	
 	\bibitem{Henneaux:1985ey}
 	M.~Henneaux, ``{ASYMPTOTICALLY ANTI-DE SITTER UNIVERSES IN D = 3, 4 AND HIGHER
 		DIMENSIONS},'' in {\em {4th Marcel Grossmann Meeting on the Recent
 			Developments of General Relativity Rome, Italy, June 17-21, 1985}},
 	pp.~959--966.
 	\newblock
 	1985.
 	\newblock
 	
 	\bibitem{henneaux1985asymptotically}
 	M.~Henneaux and C.~Teitelboim, ``Asymptotically anti-de sitter spaces,'' {\em
 		Communications in Mathematical Physics} {\bf 98} (1985), no.~3, 391--424.
 	
 	\bibitem{Ashtekar:1999jx}
 	A.~Ashtekar and S.~Das, ``{Asymptotically Anti-de Sitter space-times: Conserved
 		quantities},'' {\em Class. Quant. Grav.} {\bf 17} (2000) L17--L30,
 	\href{http://www.arXiv.org/abs/hep-th/9911230}{{\tt hep-th/9911230}}.
 	
 	\bibitem{Kapec:2015vwa}
 	D.~Kapec, V.~Lysov, S.~Pasterski, and A.~Strominger, ``{Higher-Dimensional
 		Supertranslations and Weinberg's Soft Graviton Theorem},''
 	\href{http://www.arXiv.org/abs/1502.07644}{{\tt 1502.07644}}.
 	
 	\bibitem{Hollands:2016oma}
 	S.~Hollands, A.~Ishibashi, and R.~M. Wald, ``{BMS Supertranslations and Memory
 		in Four and Higher Dimensions},'' {\em Class. Quant. Grav.} {\bf 34} (2017),
 	no.~15, 155005,
 	\href{http://www.arXiv.org/abs/1612.03290}{{\tt 1612.03290}}.
 	
 	\bibitem{Barnich:2010eb}
 	G.~Barnich and C.~Troessaert, ``{Aspects of the BMS/CFT correspondence},'' {\em
 		JHEP} {\bf 05} (2010) 062,
 	\href{http://www.arXiv.org/abs/1001.1541}{{\tt 1001.1541}}.
 	
 	\bibitem{fuks2012cohomology}
 	D.~B. Fuks, {\em Cohomology of infinite-dimensional Lie algebras}.
 	\newblock Springer Science \& Business Media, 2012.
 	
 	\bibitem{weinberg1995quantum}
 	S.~Weinberg, {\em The quantum theory of fields. Vol. 1: Foundations}.
 	\newblock Cambridge University Press, 1995.
 	
 	\bibitem{schlichenmaier2014elementary}
 	M.~Schlichenmaier, ``An elementary proof of the vanishing of the second
 	cohomology of the witt and virasoro algebra with values in the adjoint
 	module,'' in {\em Forum Mathematicum}, vol.~26, no 3, pp.~913--929, De
 	Gruyter.
 	\newblock 2014.
 	
 	\bibitem{salgado2014so}
 	P.~Salgado and S.~Salgado, ``$so (d- 1, 1)\oplus so (d- 1, 2)$ algebras and
 	gravity,'' {\em Physics Letters B} {\bf 728} (2014) 5--10.
 	
 	\bibitem{gomis2009deformations}
 	J.~Gomis, K.~Kamimura, and J.~Lukierski, ``Deformations of maxwell algebra and
 	their dynamical realizations,'' {\em Journal of High Energy Physics} {\bf
 		2009} (2009), no.~08, 039.
 	
 	\bibitem{Concha:2018zeb}
 	P.~Concha, N.~Merino, O.~Miskovic, E.~Rodr\'{\i}guez, P.~Salgado-Rebolledo, and
 	O.~Valdivia, ``{Extended asymptotic symmetries of three-dimensional gravity
 		in flat space},''
 	\href{http://www.arXiv.org/abs/1805.08834}{{\tt 1805.08834}}.
 	
 	\bibitem{Caroca:2017onr}
 	R.~Caroca, P.~Concha, E.~Rodr\'{\i}guez, and P.~Salgado-Rebolledo,
 	``{Generalizing the $\mathfrak {bms}_{3}$ and 2D-conformal algebras by
 		expanding the Virasoro algebra},'' {\em Eur. Phys. J.} {\bf C78} (2018),
 	no.~3, 262,
 	\href{http://www.arXiv.org/abs/1707.07209}{{\tt 1707.07209}}.
 	
 	\bibitem{gel1969cohomologies}
 	I.~M. Gel'fand and D.~Fuks, ``Cohomologies of {L}ie algebra of tangential
 	vector fields of a smooth manifold,'' {\em Functional Analysis and Its
 		Applications} {\bf 3} (1969), no.~3, 194--210.
 	
 	\bibitem{Unterberger:2011yya}
 	J.~Unterberger and C.~Roger, {\em {The Schr\"{o}dinger-Virasoro Algebra}}.
 	\newblock Springer, Berlin,
 	2012.
 	\newblock
 	
 	\bibitem{MR0054581}
 	G.~Hochschild and J.-P. Serre, ``Cohomology of {L}ie algebras,'' {\em Ann. of
 		Math. (2)} {\bf 57} (1953) 591--603.
 	
 	\bibitem{degrijse2009cohomology}
 	D.~Degrijse and N.~Petrosyan, ``On cohomology of split lie algebra
 	extensions,'' {\em Journal of Lie Theory} {\bf 22} (2012) 1--15,
 	\href{http://www.arXiv.org/abs/0911.0545}{{\tt 0911.0545}}.
 	
 	\bibitem{Bagchi:2010eg}
 	A.~Bagchi, ``{Correspondence between Asymptotically Flat Spacetimes and
 		Nonrelativistic Conformal Field Theories},'' {\em Phys. Rev. Lett.} {\bf 105}
 	(2010) 171601,
 	\href{http://www.arXiv.org/abs/1006.3354}{{\tt 1006.3354}}.
 	
 	\bibitem{Hartong:2015usd}
 	J.~Hartong, ``{Holographic Reconstruction of 3D Flat Space-Time},'' {\em JHEP}
 	{\bf 10} (2016) 104,
 	\href{http://www.arXiv.org/abs/1511.01387}{{\tt 1511.01387}}.
 	
 	\bibitem{Bagchi:2016bcd}
 	A.~Bagchi, R.~Basu, A.~Kakkar, and A.~Mehra, ``{Flat Holography: Aspects of the
 		dual field theory},'' {\em JHEP} {\bf 12} (2016) 147,
 	\href{http://www.arXiv.org/abs/1609.06203}{{\tt 1609.06203}}.
 	
 	\bibitem{Barnich:2015jua}
 	G.~Barnich, P.-H. Lambert, and P.~Mao, ``{Three-dimensional asymptotically flat
 		Einstein–Maxwell theory},'' {\em Class. Quant. Grav.} {\bf 32} (2015),
 	no.~24, 245001,
 	\href{http://www.arXiv.org/abs/1503.00856}{{\tt 1503.00856}}.
 	
 	\bibitem{Henkel:2017enn}
 	M.~Henkel and S.~Stoimenov, ``{Meta-conformal algebras in $d$ spatial
 		dimensions},''
 	\href{http://www.arXiv.org/abs/1711.05062}{{\tt 1711.05062}}.
 	
 	\bibitem{Barnich:2016lyg}
 	G.~Barnich and C.~Troessaert, ``{Finite BMS transformations},'' {\em JHEP} {\bf
 		03} (2016) 167,
 	\href{http://www.arXiv.org/abs/1601.04090}{{\tt 1601.04090}}.
 	
 \end{thebibliography}

\end{document}

